\documentclass[%
 reprint,
 amsmath,amssymb,
 aps,
]{revtex4-1}

\usepackage{graphicx}
\usepackage{dcolumn}
\usepackage{bm}
\usepackage[bottom]{footmisc}

\begin{document}

\preprint{APS/123-QED}

\title{Polymer translocation under a pulling force: scaling arguments and threshold forces}

\author{Timoth\'ee Menais}
\affiliation{%
CEA, INAC/SyMMES/CREAB, 17 rue des Martyrs 38054 Grenoble cedex 9 France. \\
now at UOIT, CNABLAB, 2000 Simcoe St N, Oshawa, ON L1H 7K4, Canada.
}%

\begin{abstract} DNA translocation through nanopores is one of the most promising strategies for the next-generation sequencing technologies. Most part of experimental and numerical works has focused on polymer translocation biased by electrophoresis, where a pulling force acts on the polymer within the nanopore. An alternative strategy however is emerging, which uses optical or magnetic tweezers. In this case, the pulling force is exerted directly at one end of the polymer, which strongly modifies the translocation process. In this paper, we report numerical simulations of both linear and structured (mimicking DNA) polymer models, simple enough to allow for a statistical treatment of the pore structure effects on the translocation time probability distributions. Based on extremely extended computer simulation data, we : {\em i)} propose scaling arguments for an extension of the predicted translocation times $\tau \sim N^{2}F^{-1}$ over the moderate forces range; {\em ii)} analyze the effect of pore size and polymer structuration on translocation times $\tau$.

\end{abstract}

\maketitle


\section{Introduction}
\label{sect:introduction}
The translocation of biomolecules through nanopores has been an active research field for more than two decades now, implying both experimental and theoretical interesting aspects. Translocation consists in a biomolecule crossing a membrane through a hole of nanometric size (the nanopore) from the cis- to the trans- side. This process can either be natural (unbiased) or forced. Two main methods have emerged in the latter case. In the first, the translocating polymer is driven by an electrophoretic field by applying a potential bias between cis and trans sides. In the second, driving consists of a mechanical forces applied directly to one end of the polymer by using optical or magnetic tweezers. Kasianowicz {\em et al.}~\cite{Kasianowicz1996} were first to successfully perform a forced DNA translocation through a biological nanopore. This experiment paved the way for the use of translocation to characterize single objects such as single or double stranded DNA, proteins, or, at larger scales, even cells~\cite{Dekker2007, Keyser, Meller2003, Movileanu2008, Palyulin2014,Wanunu2012, Liu2013}. High impact applications of translocation in biotechnology and medical diagnostics are expected, with clear emphasis on quicker and cheaper methods for DNA sequencing~\cite{Branton2008,Clarke2009,Merchant2010,Schneider2012}.

Current limitations to the use of DNA translocation as a sequencing tool are both spatial and temporal. More precisely, in the case of common biological and artificial nano-pores, several nucleotides are present simultaneously within the nanopore during the translocation. This hinders the possibility of single-base sequence resolution~\cite{Schneider2010}. Also, in most experiments a base spends about 1~$\mu$s within the pore, while current measurement resolution times would require a significant slowing down of the process, with a typical occupation time of about 1~ms~\cite{Branton2008}. Although attaching an optical (or magnetic) bead at the extremity of a polymer is substantially less straightforward compared to direct electrophoresis, this method is an effective candidate for a better control over the translocation process. In particular,  it has been shown that it can reduce the mean translocation time by one order of magnitude~\cite{Kantor2004}. Another potentially valuable strategy to reduce the translocation speed is to use {\em narrow} nano-pores thus adding supplementary friction to the system. Both the pulled translocation~\cite{Keyser2006, Peng2009, VanDorp2009, Sischka2010, Bulushev2014, Bulushev2015, Bulushev2016} and the effects of the nanopore size~\cite{Yameen2009} have been investigated in experiments. Only few theoretical or numerical studies, however, have been carried out for both cases~\cite{Kantor2004, Grosberg2006, Huopaniemi2007, Panja2008, deHaan2010, deHaan2015, Menais2016,2017arXiv170809184S}. Furthermore, contrary to the case of a field driven translocation that has been extensively investigated~\cite{Sakaue2007,Sakaue2010,Saito2011,Saito2012,Ambjrnsson2005,2Ambjrnsson2005,Rowghanian2011,Dubbeldam2012,Ikonen2012,2Ikonen2012,Ikonen2013}, the main physical understanding of the translocation process of pulled polymers through nanopores remains controversial, only early results are available for high forces~\cite{2017arXiv170809184S}. This is mainly based on the methods of statistical mechanics, looking for universal mechanisms controlled by general parameters. In particular, there are discrepancies about the correct values for the critical exponents $\alpha$ and $\delta$ governing the translocation time, $\tau \propto N^{\alpha}F^{-\delta}$, as function of the polymer length $N$ and the applied force $F$~\cite{Milchev2011}.
\begin{figure*}[t]
\centering
\includegraphics[width=0.4\textwidth,angle=90]{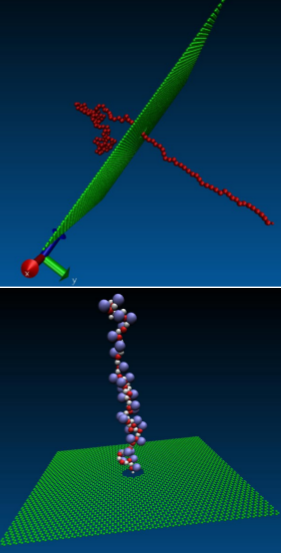}
\caption{Polymer during a translocation process through the nanopore aperture of the membrane. Membrane beads are in green. Left: Linear polymer grains are depicted in red. Right: Structured polymer, the backbone is in red (sugar beads) and white (phosphate beads) while the lateral grains in purple are bases bonded to the sugar beads (red) from the backbone.}
\label{FigModel}
\end{figure*}

In this paper, we report numerical simulations of a simple model for linear and structured polymers, translocating through a nanometric pore carved in a structured immobile membrane. We have considered toy numerical models which allow for a complete statistical treatment of the translocation processe. The {\em linear} polymer is a classic Rouse-like chain formed by connected beads. The {\em structured} polymer is obtained by the same linear chain (representing the polymer backbone), with additional beads regularly grafted to the backbone and mimicking single-stranded-DNA bases, at the coarse-grained level. The membrane is constituted by immobile interaction centers arranged on a hexagonal lattice reminiscent of the structure of graphene.  Based on extended sets of data, we provide general insight on the effect of both polymer structure and nano-pore sizes. We propose simple scaling arguments based on the waiting times inside the pore to confirm the predicted value for $\alpha = 2$~\cite{Huopaniemi2007, Panja2008} for intermediate range forces and expand recent results~\cite{2017arXiv170809184S} to this regime. Furthermore, we demonstrate the existence of a threshold force for small pore size and we relate the underestimation of the predicted value $\delta=1$~\cite{Huopaniemi2007, Panja2008} at high forces by numerical simulations to the bond stretching~\cite{Luo2009}.
\begin{figure*}[t]
\centering
\resizebox{0.3\textwidth}{!}{\includegraphics[trim={2cm 0 2cm 0 0}]{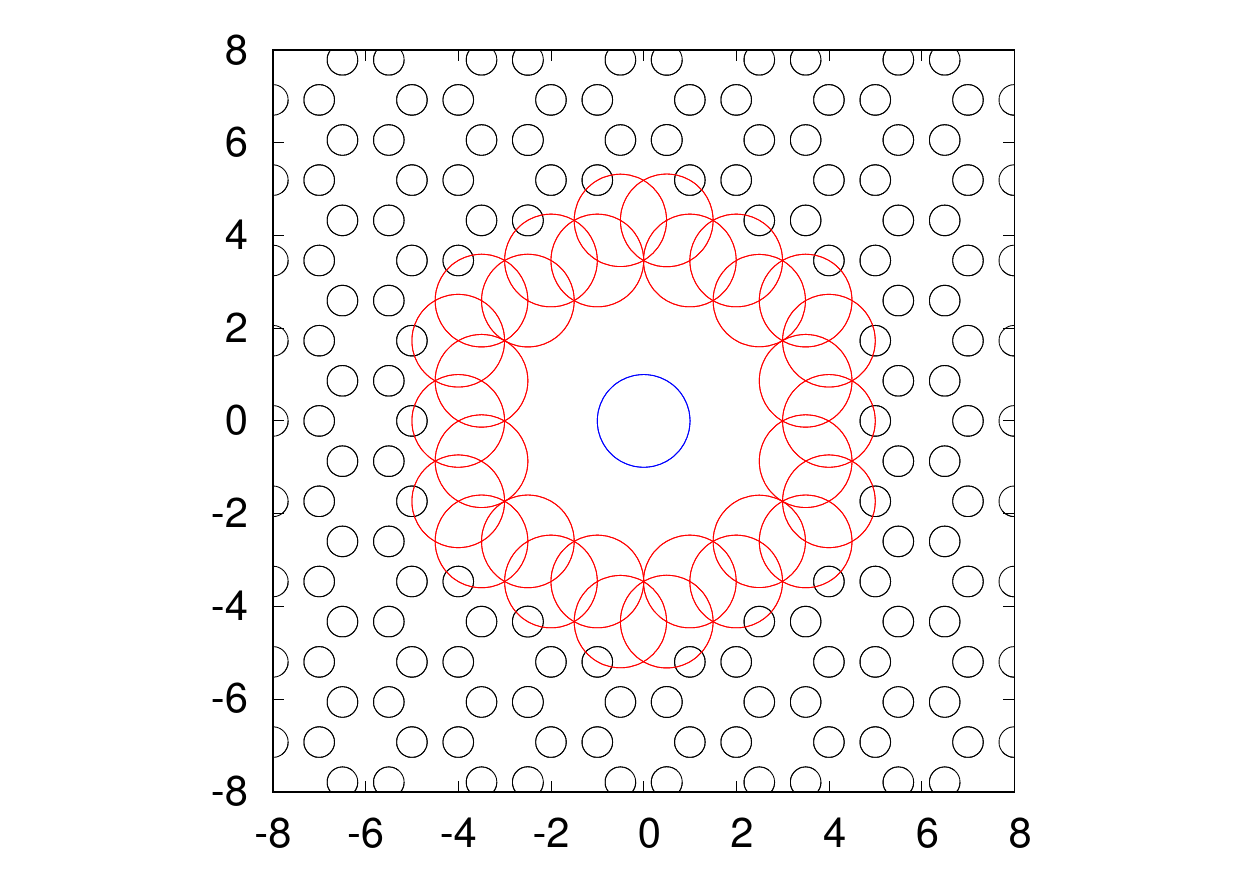}}\resizebox{0.3\textwidth}{!}{\includegraphics[trim={2cm 0 2cm 0 0}]{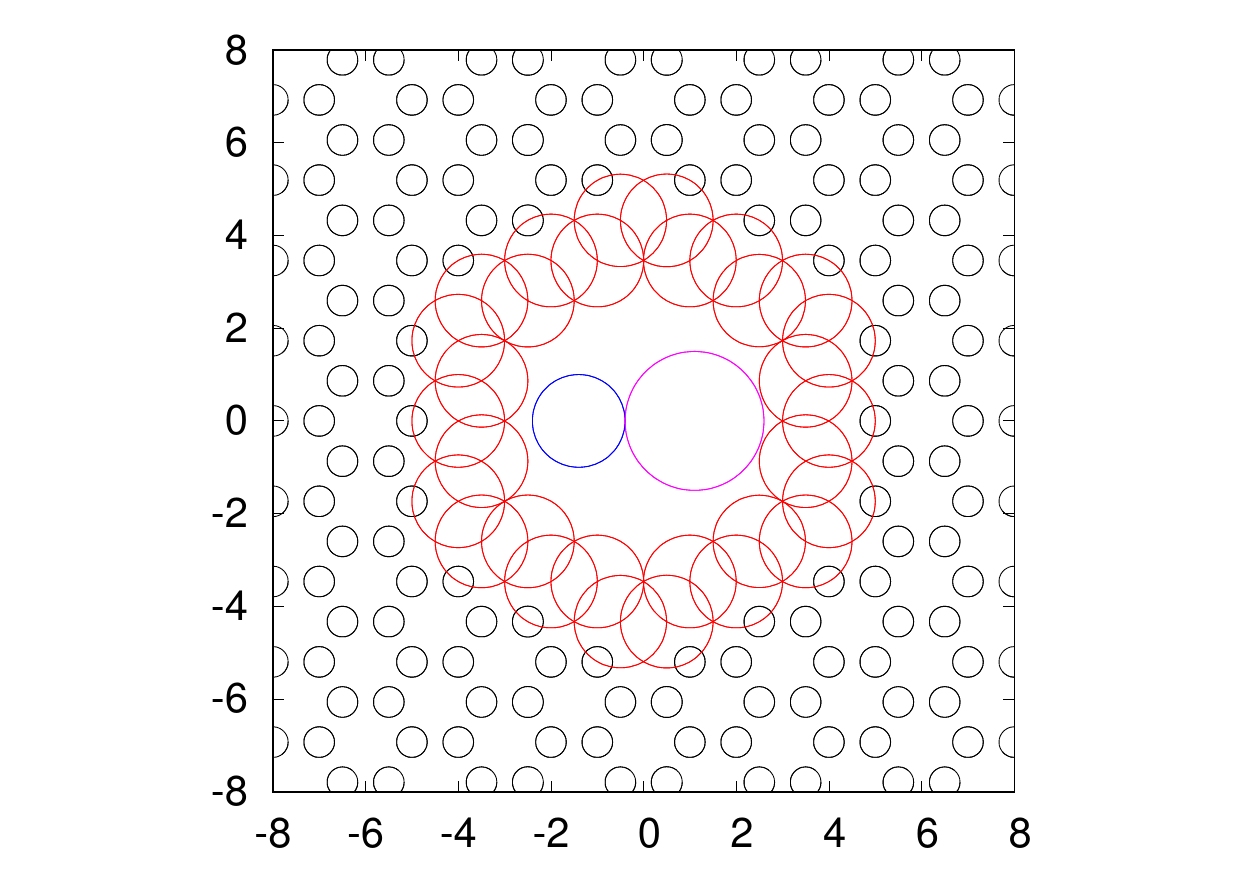}}\resizebox{0.3\textwidth}{!}{\includegraphics[trim={2cm 0 2cm 0 0}]{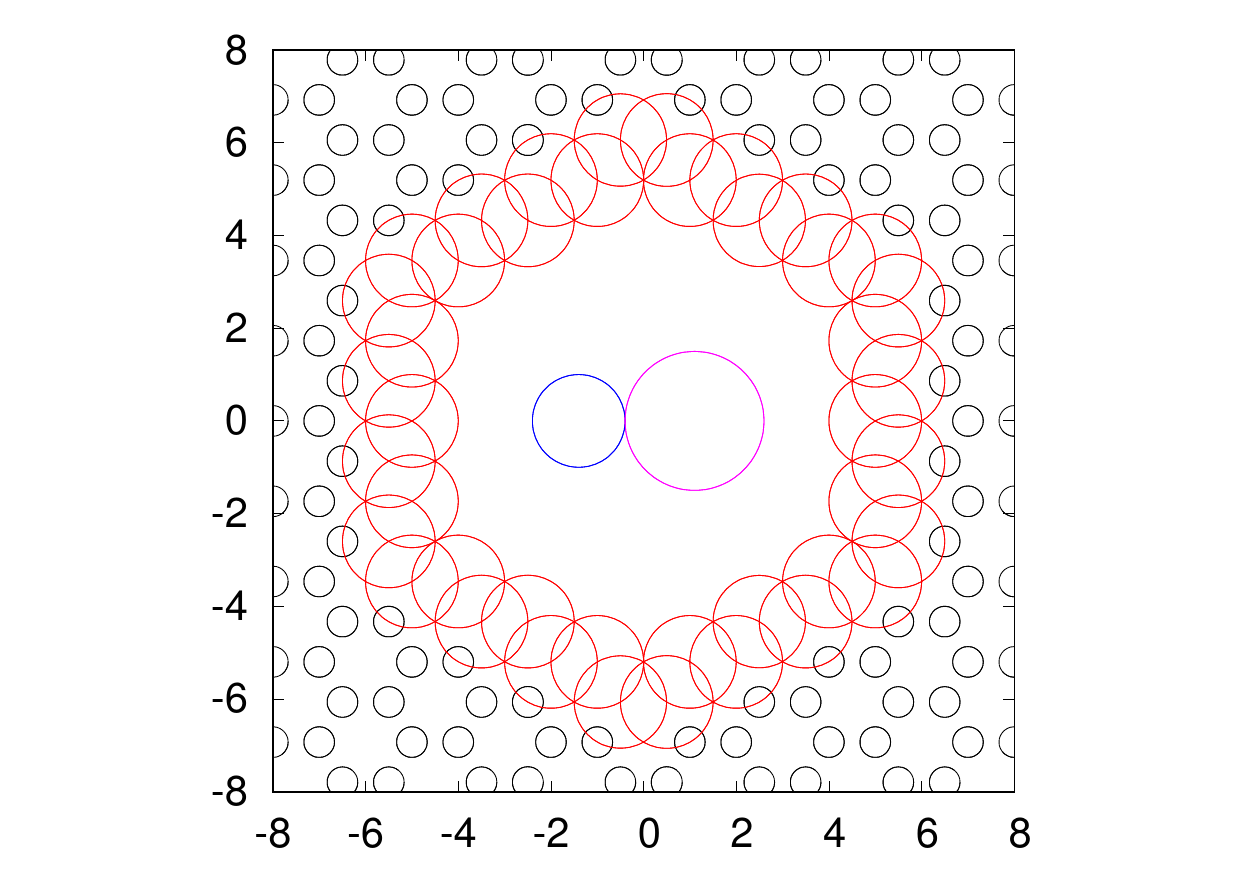}}\rotatebox{90}{\resizebox{0.33\textwidth}{!}{\hspace{-0.8cm}\includegraphics[height=0.07\textwidth]{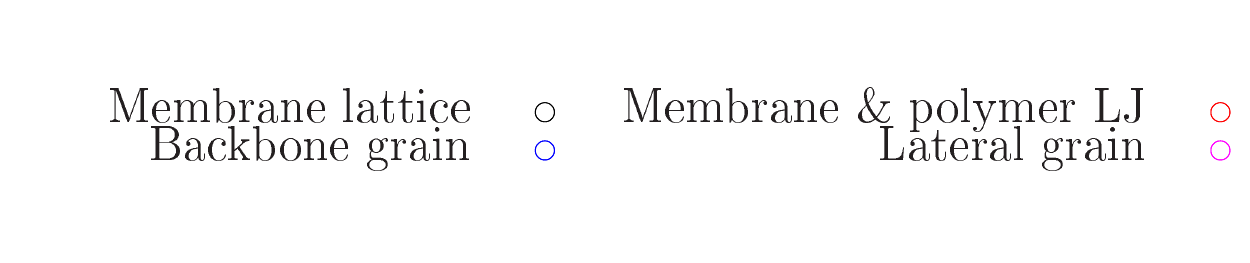}}
}
\caption{Representation of the various pore sizes studied with the polymer beads represented inside the pore. Backbone beads are depicted in blue and lateral bases in purple with their respective LJ radius. The membrane beads are depicted in black except the first layer of the nanopore depicted in red with their respective LJ radius for clarity. (Left) Large pore size for a simple linear polymer. (Middle) The same pore size corresponds to a small pore size for the structured polymer. (Right) Large pore size for the structured polymer.}
\label{FigPlotofpores}
\end{figure*}
\section{Models and simulation details}
\label{sect:methods}
In view of a complete statistical analysis of the translocation time and its dependence on the polymer structure and nanopore size, we have considered a coarse-grained model for both the polymer and the membrane. Our aim is to define a model simple enough to produce very extended ensembles of translocation events in order to completely characterize the statistical properties of the translocation time. However, we aim at keeping the essential properties of the DNA translocation which means to consider also a structured polymer model with additional bases attached to the backbone. We give below a few details about the models and the molecular dynamics simulation procedures. We refer the reader to the Supplemental Material (SM) \footnotemark[1] \footnotetext[1]{See Supplemental Material at [URL will be inserted by publisher] for supporting text and figures} and Ref.~\cite{Menais2016} for additional details. 
\subsection{The linear and structured polymer models}
A minimalistic model for single stranded DNA was considered incorporating three bead types~\cite{Menais2016}. The alternating phosphate (P) and sugar (S) beads form the linear polymer backbone (Fig.~\ref{FigModel}, Top). Lateral beads, modeling the DNA bases (B), are attached to the S-beads of the polymer backbone to form the structured polymer (Fig.~\ref{FigModel}, Bottom). Note that no distinction is made at this level between the four bases (A-C-T-G). The total number of chain beads is $N$, one supplementary bead represents the pulling bead. For the structured polymer a lateral grain is attached every even chain bead ($N/2$ lateral beads). 

Steric and binding interactions only have been incorporated in the model. Steric interactions between any two beads, $i$ and $j$, are modeled by a truncated and shifted Lennard-Jones (LJ) potential:
\begin{equation}
U_{LJ}(r_{ij})= 4\epsilon \left[\left(\frac{\sigma_{ij}}{r_{ij}}\right)^{12}-
\left(\frac{\sigma_{ij}}{r_{ij}}\right)^{6}\right] + \epsilon
\label{Eq:LJ}
\end{equation}

with $U_{LJ}(r_{ij})=0$ for $r_{ij}\le 2^{1/6} \sigma_{ij}$. Here  $r_{ij}$ is the distance between $i$ and $j$, $\epsilon_{ij}=\epsilon=1$, and $\sigma_{ij} = (\sigma_i + \sigma_j)/2$ with $\sigma_i$ the radius of bead $i$. This potential is therefore purely repulsive and prevents the beads from interpenetrating at very short distances. The sizes of the polymer backbone beads are $\sigma_{S}=\sigma_{P}=a=0.3$~nm. The size of the lateral bases in the structured polymer are slightly larger, $\sigma_{B}=3/2 a=0.45$~nm. Bonds between adjacent beads (both pertaining or grafted to the backbone) are modeled by the Finitely Extensible Nonlinear Elastic (FENE) anharmonic potential:
\begin{equation}
U_{FENE}(r_{ij})= -15\epsilon \left(\frac{R_{ij}}{\sigma_{ij}}\right)^2  
\ln\left[1-\left(\frac{r_{ij}}{R_{ij}}\right)^2\right]
\label{Eq:fene}
\end{equation}
with the the bond length value, $R_{ij}=3/2\sigma_{ij}$. 

Hydrogen bonding, base stacking and backbone bending are, therefore, not taken into account and, as a consequence, we cannot tackle issues involving the effect of stiffness, or the helicoidal structure building of the DNA double helix~\cite{Linak2011}. Note however that, since we focus our study on short single stranded DNA translocation without secondary structures formation like hairpins, these omitted degrees of freedom are expected to be irrelevant in the present context. To further reduce computational costs allowing for an accurate an statistical sampling of the translocation process, we disregard the presence of (explicit) solvent and neglect hydrodynamic interactions. Only Rouse dynamics of polymers is thus considered.

We have tested the consistency of the considered polymer models with the exact predictions for the free polymer chain (see Sect.~II of the SM \footnotemark[1]). More in details, we have verified that the diffusion constant scales as the inverse of the polymer length, $D\propto N^{-1}$, due to the absence of hydrodynamic interactions. The friction of the polymer also scales as the length of the polymer, and we have also verified that the fluctuation-dissipation relation relating the two variables  is fulfilled. Finally, the polymer radius of gyration (similar to the end-to-end distance) scales as $N^\nu$. The Flory exponent $\nu \simeq0.58$ is close to $\nu = 0.588$ expected for a Rouse-like linear polymer, while a higher $\nu \simeq 0.71$ has been measured for the structured polymer. These latter result can be rationalized in terms of an effective  larger persistence length due to the presence of the side beads, and thus of stronger finite size effects overestimating the exponent $\nu$.

For translocation purposes a constant force is exerted at the incoming end 
of the polymer as performed experimentally with optical or magnetic 
tweezers~\cite{Bell2012, HernndezAinsa2013, Peng2009, Trepagnier2007,VanDorp2009}. 
\subsection{The membrane}
To remove effects originating from a non-negligible thickness of the nano-pore, we consider a perfectly 2-dimensional membrane formed by carbon (C) atoms arranged on a honeycomb lattice (see Fig.~\ref{FigPlotofpores}) reminiscent of the graphene structure. In our model, we also preserve a realistic ratio between the typical length scales associated to graphene and DNA, respectively. This amounts to a membrane lattice constant $b=a/2=0.15$~nm. Periodic boundary conditions are used in the plane parallel to the membrane, while free boundaries are employed in the direction orthogonal to the pore. We note that the size of the membrane is chosen sufficiently large to accomodate a fully stretched polymer while avoiding self-interactions with its own image. The membrane grains are immobile, and interact with the polymer beads via Eq.~(\ref{Eq:LJ}) with $\sigma_C=1$, ensuring that polymer beads cannot cross the membrane outside the nano-pore. In Fig.~\ref{FigModel} we show typical snapshots (generated by VMD~\cite{Humphrey1996}) of both polymer models pulled through the nanopore carved in the membrane. Details of the color code used are given in the caption of the figure. Figure~\ref{FigPlotofpores} represents the different sizes of polymer grains (blue and pink) with respect to the pore through which their translocation is studied (red).
\subsection{Molecular Dynamics Simulations}
Concerning the time integration, an efficient parallelised algorithm is used by the LAMMPS~\cite{Plimpton1995} software to solve the Langevin equation of motion
\begin{equation}
\large{m_n \frac{\partial^2 \textbf{r}_n}{\partial t^2} = 
-\frac{\partial U_{n}}{\partial \textbf{r}_n} - \nu_n \textbf{v}_n   
+ \textbf{g}_n}
\label{Eq:lang1}
\end{equation}
with $m_n$ and $\textbf{r}_n$ being the mass and the position of the $n^{th}$ grain, $t$ the time, $U_{n}$ the sum of all the potentials applied to the $n^{th}$ grain, $\nu_n$ the friction coefficient of the $n^{th}$ grain and $\textbf{g}_n$ is the white thermal noise, of average $\left<\textbf{g}_n\right> = 0$, and variance $\left<\textbf{g}_n^2\right>=6\nu_nk_B T$. The temperature $T$ of the polymer beads is set to $3/2\text{ }\epsilon$. In the following both the mass and the friction coefficient are considered independent of the beads $m_n=m$ and $\nu_n=\nu$. Eq.~\ref{Eq:lang1} is integrated numerically with a time step $\delta t=5 \text{ } 10^{-3}$ ($26$ fs in SI units).

Translocation simulations started with an equilibrated polymer with one end attached at the entrance of the nanopore. For each forces, a set of 1000 translocations were performed for statistical analysis. We have used absorbing boundary conditions (i.e. if the first monomer does not cross the membrane and wanders in the cis side, the simulation is aborted). 
\section{Results and discussion}
\subsection{Linear polymer}
We considered as a reference polymer, the general case of Rouse-like linear polymer 
in order to be able to compare with the case of the structured polymer.
The translocations have been performed through a large pore by pulling one end of 
the polymer (see top of Figure~\ref{FigModel}). The translocation of polymer
forced by electrostatic interactions imposed by a voltage difference between 
the cis and trans side of the membrane have been extensively studied in the literature.
However, the case of end polymer pulling while experimentally relevent~\cite{Keyser2006, 
Peng2009, VanDorp2009, Sischka2010, Bulushev2014, Bulushev2015, Bulushev2016} 
has been poorly considered by numerical simulations~\cite{Kantor2004, Huopaniemi2007,
Panja2008} until very recently~\cite{2017arXiv170809184S}.

\begin{figure}[t]
\begin{center}
\includegraphics[width=0.5\textwidth]{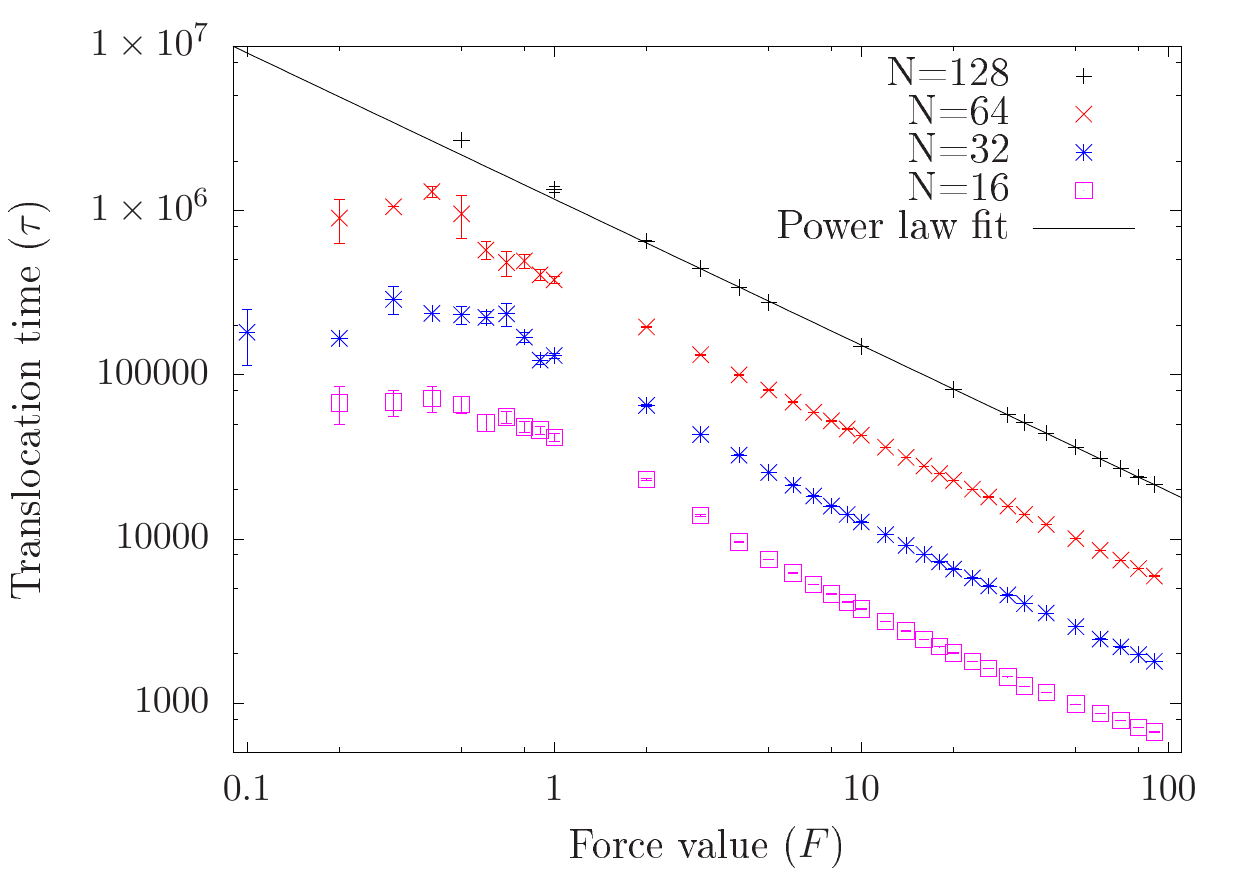}
\includegraphics[width=0.5\textwidth]{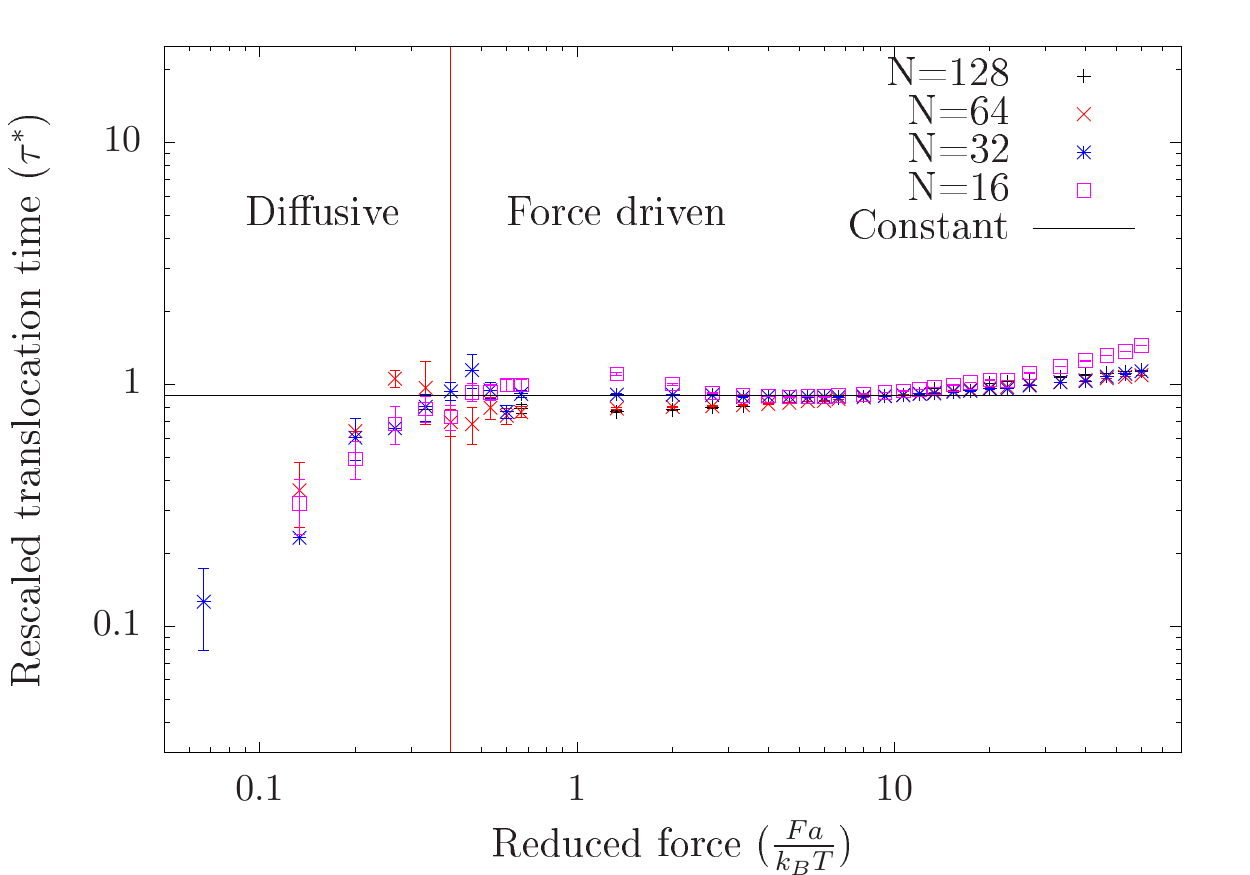}
\caption{Mean translocation time $\tau$ (top) and its rescaled value $\tau^*$ (bottom) 
as function of the bias pulling force (top) and the reduced force $Fa/k_BT$ (bottom). 
We investigate four polymerisation degrees ($N=16,32,64,128$ respectively in pink, 
blue, red and black) over three orders of magnitude for the bias force (from 0.1 to 100). 
The rescaled mean translocation time is as $\tau^*=\tau*F/N^{1.78}$. 
The rescaling of the translocation time clearly shows two regimes corresponding 
to a diffusive regime translocation and a force driven one.}
\label{FigLinpolTranslocationtime}
\end{center}
\end{figure}

Theory for biased polymer translocation aims at finding so called critical exponents, 
$\alpha$ and $\delta$, linking the mean translocation time to namely the polymerisation 
index and the bias force: $\tau \propto N^\alpha / F^\delta$. As formerly discussed, 
many theories and simulations investigate the case of a gradient bias force applied 
in the center of the nanopore~\cite{Sakaue2007,Sakaue2010,Saito2011,Saito2012,Ambjrnsson2005,2Ambjrnsson2005,Rowghanian2011,Dubbeldam2012,Ikonen2012,2Ikonen2012,Ikonen2013}. In this study, we are interested in the case of a 
pulling force at one end of the translocating polymer. Kantor and Kardar were 
the first to tackle the issue of a translocation driven by a pulling force and 
predicted  $\alpha=2$ and $\delta=1$ scaling exponents~\cite{Kantor2004}. 
Their prediction is backed by 2D MD simulations which lead to a lower critical 
exponent $\alpha=1.875$~\cite{Kantor2004}. Such values were later confirmed in the 
literature~\cite{Huopaniemi2007} and are in agreement with this present work 
(see Figure~\ref{FigLinpolTranslocationtime}).

We find that the mean translocation time shows two regimes as expected from the 
literature~\cite{Kantor2004,Huopaniemi2007}. In both regimes, $\alpha = 1.78$, only the value of $\delta$ differs. 
For low pulling forces, we have a diffusive regime ($\delta=0$) and a force driven
regime when the force is increased ($\delta=1$). For very high forces we observe  
a decrease in the critical exponent (increasing behaviour of the rescaled 
translocation time). The underestimation of the critical exponent $\delta \simeq 
0.8 < 1$ at high forces is due to bond strengths and/or friction coefficients 
set too low. Those findings were also observed when the bias force originates 
from a difference of potential~\cite{Luo2009}. We analyzed deterministic
polymers at zero temperature for various frictions and bond strengths (see
Supplementary Information Section II). Our findings show the importance 
of such parameters in the value of the critical exponent $\delta$. 
Furthermore, in the precise case of a pulled polymer, the crowding in the trans 
side may not be responsible for this decrease in $\delta$ as suggested by one of the last reviews on the topic~\cite{Palyulin2014} 
(the crowding in the trans side does not occur for a polymer under traction see Figure~\ref{FigModel} and SM Section II \footnotemark[1]). Those results are in agreement with recent results for field driven translocations that suggest the trans side crowding is irrelevant and force tension (directly connected to the parameters fore mentioned) alone can explain the decrease of the critical exponent $\delta$~\cite{2017arXiv170706663S}.

Looking at the translocation time distribution (see SM
Section III \footnotemark[1]), we find that they are in agreement with a first passage 
probability density function proposed by  Ling and Ling~\cite{Ling2013} 
for a translocation biased by electrophoresis and confirmed for the case 
of a pulling force~\cite{deHaan2015}. 
The fitting parameters are the diffusion coefficient of the membrane 
along the polymer and the mean translocation velocity. An accurate 
estimation of the diffusion coefficient is impossible due to a large 
range of validity for the parameter. However, one can estimate the mean 
translocation velocity which coincides with the average speed deduced from
the mean translocation time divided by the chain length.
\begin{figure}[t]
\begin{center}
\includegraphics[width=0.5\textwidth]{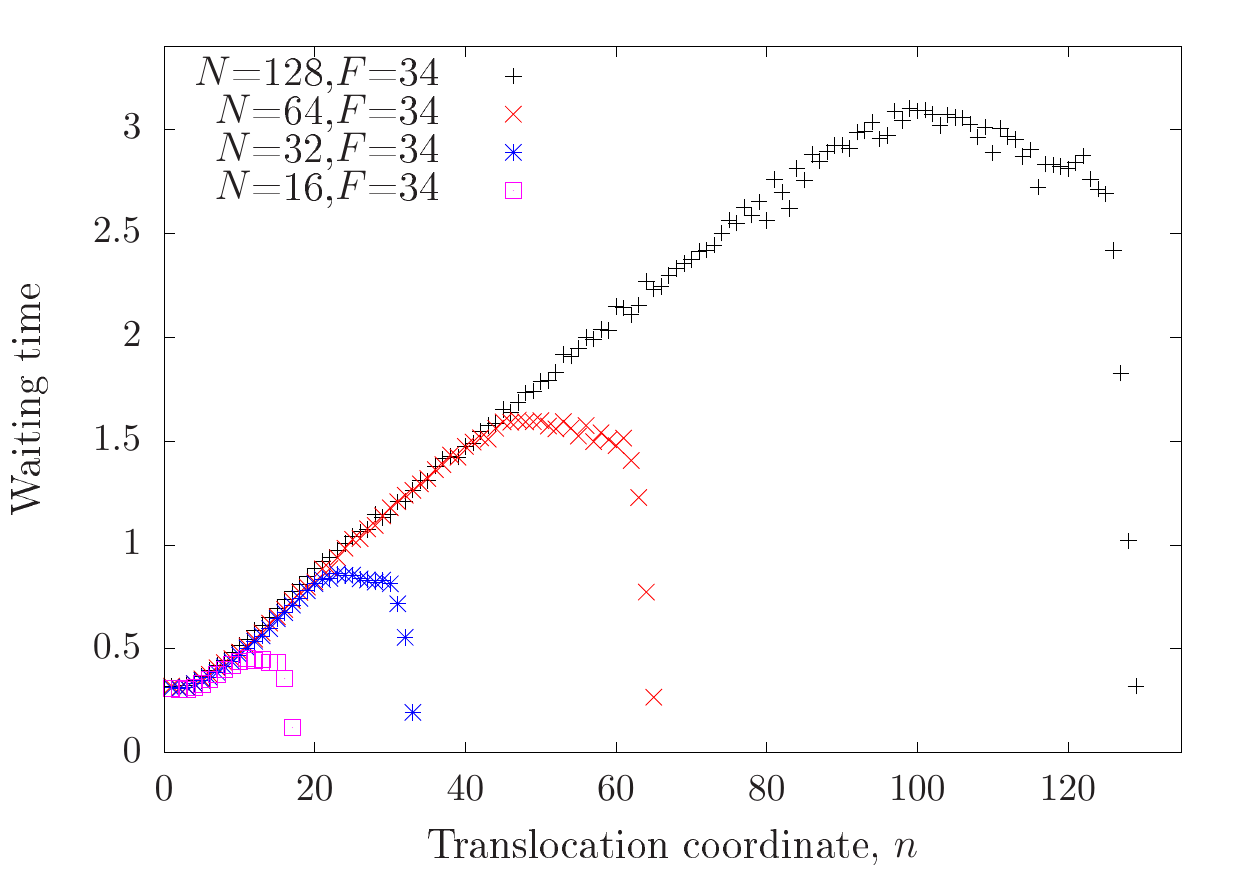}
\caption{Average waiting time depending on the translocation coordinate 
at a given force ($F=34$) for 4 polymer lengths ($N=16, 32, 64, 128$, 
respectively in pink, blue, red and black).  The average waiting time 
curve presents the same pattern composed of 3 different regimes for 
the 4 polymer lengths studied: a linear regime dominated by the pulling force,
then a plateau dominated by diffusion and finally a collapse of the waiting time 
due to tail retraction from the cis side.}
\label{FigLinpolwaitingtime}
\end{center}
\end{figure}

We pushed the analysis further by looking at the average waiting time for 
the translocation coordinate at different polymerisation degrees. 
The translocation coordinate corresponds to the number $n$ of monomers already 
translocated to the trans side. The average waiting time is the average time 
that the $n^{th}$ monomer needs to translocate (including possible returns 
if $n$ is not equal to 1 or $N$). The sum of all the average waiting time over 
the polymer length gives the average mean translocation time. We find 
waiting times curves in agreement with the general shape already published 
in the literature~\cite{Huopaniemi2007} (see Figure~\ref{FigLinpolwaitingtime}). 
In particular, the average waiting time as function of the translocation coordinate 
behaves linearly at the beginning of the translocation and for a given pulling force. 
Thus, at the beginning of the process, the translocation is entirely driven by 
the pulling force whose influence linearly decreases as it is spread on the $n$ 
monomers already translocated. The force is felt only by the trans side and the 
linear regime is present as long as the diffusion is negligible. When the diffusion 
force from the cis side compares to the pulling force, a plateau regime is reached. 
This second regime ends sharply with the retraction of the polymer tail from 
the cis side. This collapse or retraction of the polymer tail occurs over a 
constant value of 4-5 monomers.

\begin{figure}[t]
\begin{center}
\includegraphics[width=0.5\textwidth]{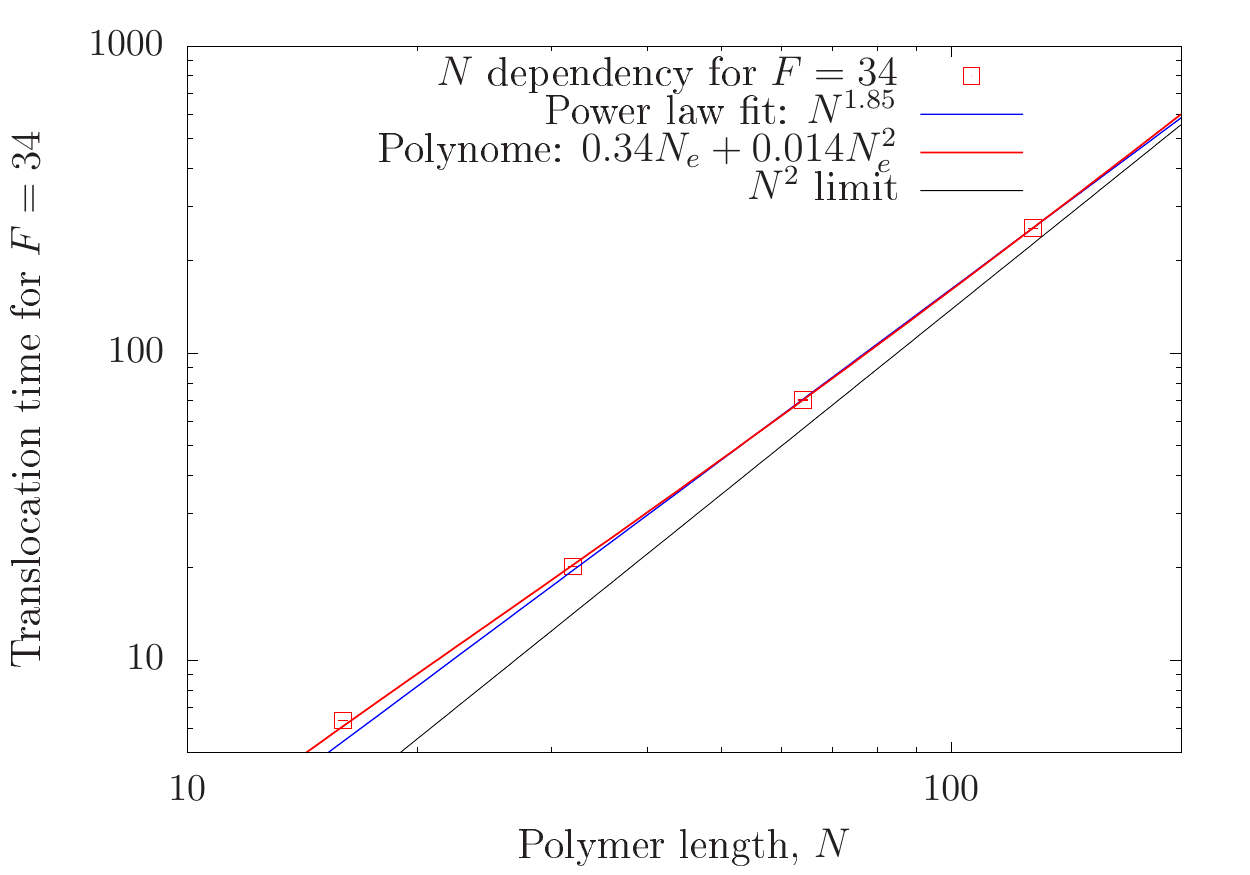}
\caption{Mean translocation time as function of the polymer length $N$ 
at a given pulling force $F=34$. A second order polynomial fits 
the numerical results with better accuracy than a simple power law
with critical exponent $\alpha = 1.85 < 2$ provided one takes into 
account the effect of the polymer tail retraction ($N_e = N-4$).}
\label{FigScaleresult}
\end{center}
\end{figure}

From those observations, we develop a scaling analysis of the translocation 
of pulled polymers. From Figure~\ref{FigLinpolwaitingtime}, the waiting 
time $w(n) = a(F) + b(F) n$ is linear for the force driven regime ($n<\beta N$) 
with a constant term $a(F)$ and a slope $b(F)$. The second unbiased regime 
corresponds to a quasi-constant waiting time although slightly decreasing 
when $(N-n) \gg 1$ and sharply decreasing on 4-5 monomers for the third regime 
of polymer tail retraction. Thus, we obtain the following mean translocation time:
\begin{eqnarray}
\tau(F,N) & = & \int_0^{\beta N} dn\; [a(F)+b(F)\;n] \nonumber \\
& + &\int_{\beta N}^{N} dn\; [a(F) + b(F)\;\beta N]. 
\label{EqIntegral}
\end{eqnarray}
%
Numerical evidence shows that $a(F)$ and $b(F)$ are inversely proportional to $F$, and $\beta N$ (or at least the arithmetical prefactor) remains constant on a large range of parameters (see calculation in appendix and SM Section IV \footnotemark[1]), hence suggesting the 
following scaling for the translocation time:
%
\begin{eqnarray}
\tau(F,N) = \Gamma \frac{N^2}{F} + \frac{{\cal O}(N)}{F}
\label{EqTauformula}
\end{eqnarray}
where $\Gamma$ is a constant. 

A second order polynomial fit of the translocation time as function of the 
polymer length using $N_e = N-4$ to avoid the 4-5 tail retracting monomers 
was adjusted to the numerical results (See Figure~\ref{FigScaleresult}). 
The agreement obtained is clearly better than a simple power law scaling 
with exponent $\alpha = 1.85$. The next to leading behavior term in (\ref{EqTauformula}) is similar to what was found by Ikonen et al.~\cite{Ikonen2013} in the case of gradient biased translocation with $\tau \propto N^{1+\nu} + O(N)$. $N^{1+\nu}$ is the asymptotic value and the $O(N)$ term is due to pore friction. For intermediate pulling forces, friction from the trans side and the pore are the main contributors to explain the scaling of the mean translocation time.

The idea that friction from the trans side can have an important role was first mentioned in the case of a semi flexible polymer for an electric field driven translocation~\cite{Sarabadani2017}. For a very high pulling force recent results~\cite{2017arXiv170809184S} show that one has to take into account both friction from the trans side and from the cis side due to tension propagation. Those results stand as long as the force is high enough so that the polymer would always be fully stretched if pulled in the bulk. Based on work conducted on polymers in the bulk~\cite{Sakaue2012}, the minimum force should satisfy $F> 3k_BTN/a$ which would correspond to unit-less forces greater than 300 in most simulations and is questionably satisfied in ref~\cite{2017arXiv170809184S}.

Let us predict an intermediate forces scaling based on our simulations and theoretical work from Sarabadani et al.~\cite{2017arXiv170809184S} that was conducted for high pulling forces and yields:

\begin{eqnarray}
\tau(F,N) =   \frac{1}{F} \left[\tilde{\eta}_PN + \frac{A_\nu}{1+\nu}N^{1+\nu} + \frac{1}{2}N^2 \right]
\label{EqTauformulaforhighpulling}
\end{eqnarray}
With $\tilde{\eta}_P$ the pore friction coefficient and $A_\nu$ the prefactor in the scaling of the end to end distance of the polymer. The three terms of eq.~\ref{EqTauformulaforhighpulling} are the respective friction contributions of the polymer with the pore, the cis side and the trans side. In our polynomial fit using eq.~\ref{EqTauformula} the prefactor of the $N^2$ contribution is in agreement with the prefactor of eq.~\ref{EqTauformulaforhighpulling} and pore friction contribution remains relevant. It seems that the contribution of the cis side friction is weak when the force is intermediate.
Our reasoning is the following: in the case of a high pulling force, eq.~\ref{EqTauformulaforhighpulling} stands but if $F<3Nk_BT/a$, as soon as there is blob formation, the pore prevents the trumpet formation, the tension in the cis side vanishes and so does it's influence on the mean translocation time. The mean translocation time scaling becomes:

\begin{eqnarray}
\tau(F,N) =   \frac{1}{F} \left[\tilde{\eta}_PN + \frac{A_\nu}{1+\nu}\left(\frac{aF}{3k_BT}\right)^{1+\nu} + \frac{1}{2}N^2 \right]
\label{EqTauformulaintermediate}
\end{eqnarray}
For $N>aF/3k_BT$ ; for smaller chains, one recovers the high forces limit scaling.

 When the waiting time behaves linearly, it is clearly a $N^2$ contribution to the mean translocation time which indicates trans side friction is predominant. The contribution of the cis side friction is mainly seen in ref~\citep{2017arXiv170809184S} at the begining of the translocation when the tension propagation towards the cis side is maximum. This contribution with a mean waiting time increasing slower than linearly can also be seen in the early stages of the translocation on Figure~\ref{FigLinpolwaitingtime}. The fact that waiting times are independent of the chain length at the begining of the translocation process is further indication that the cis side friction contribution is constant and becomes negligible with the increase of the chain length in the intermediate pulling forces range.

\subsection{Structured polymer translocating through large and narrow pores}

The linear polymer discussed previously will now serve as a reference 
for the translocation of structured polymers through narrow and large  
pores (see narrow and large pore sizes on Figure~\ref{FigPlotofpores} 
middle and bottom respectively).

The general behaviour of the translocation timescale observed for the 
linear polymer remains for the structured polymerwhen translocating 
throug the large pore. The difference is a $3/2$ rescaling factor of 
the timescale due to the larger number of beads present in the 
structured polymer. The proportion of beads between the structured 
and linear polymer is $3/2$ due to a side bead corresponding to the 
bases attached every two beads on the backbone. 
Furthermore, for the structured polymer, the crossover between unbiased and 
biased regimes is not as clear as previously observed for the linear polymer.  
Although, this does not seem to affect the critical exponents $\alpha = 1.76$ 
and $\delta \simeq 1$.
\begin{figure}[t]
\begin{center}
\includegraphics[width=0.5\textwidth]{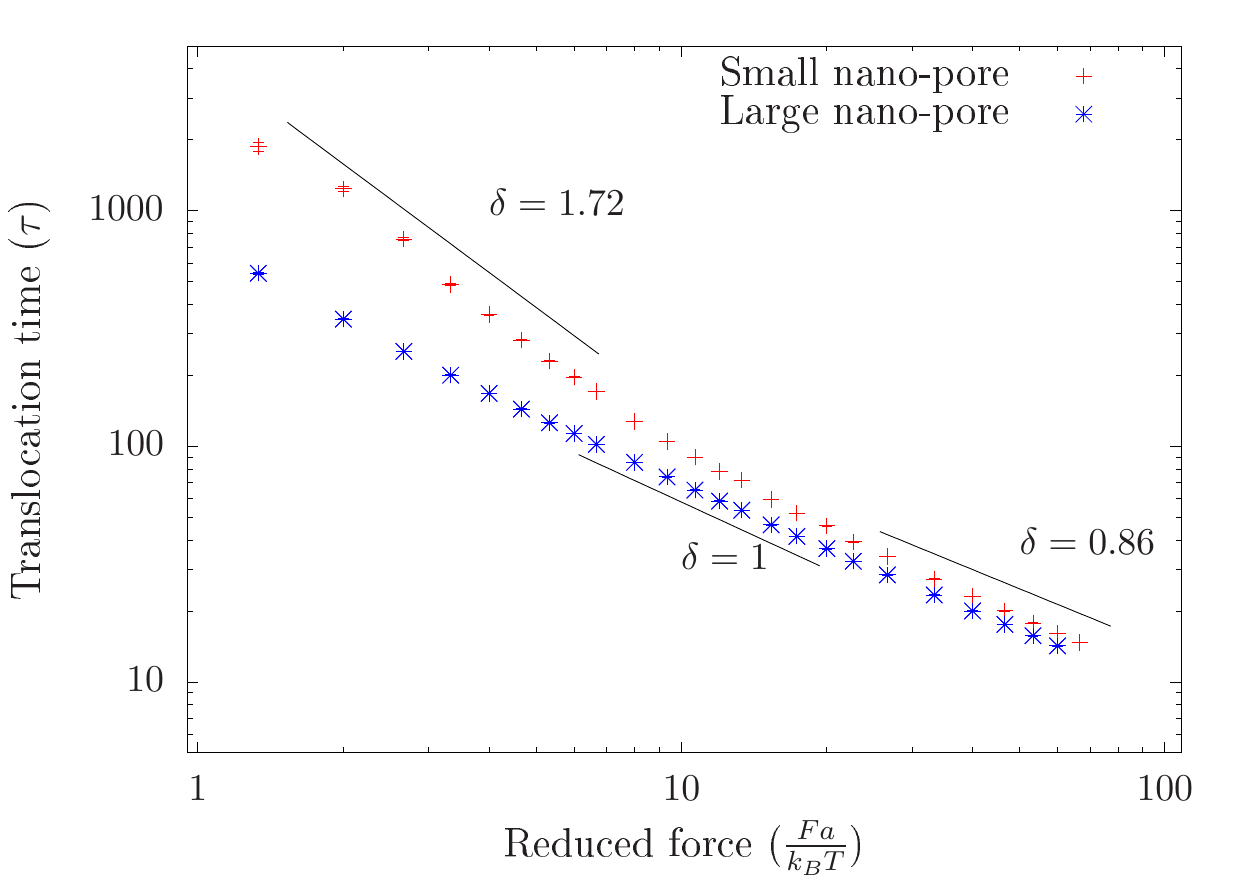}
\caption{Mean translocation time for polymer chain length of $N=32$ for 
a large and a narrow pore (reproduced from our previous work~\cite{Menais2016}). 
At high pulling forces, their behaviour is 
similar with $\delta<1$ due to bonds elongation. When the bias force is 
decreased, a sharp increase in the translocation for the narrow pore 
compared to the large one occurs. This difference can be up to one order 
of magnitude and when the bias force is too low, the process does not 
even occur (hence the investigation of only 2 orders of magnitude in 
force range for the narrow pore). No general $\alpha$ and $\delta$ value 
can be established in the moderate forces regime.}
\label{FigComppores}
\end{center}
\end{figure}

\begin{figure}[t]
\begin{center}
\includegraphics[width=0.5\textwidth]{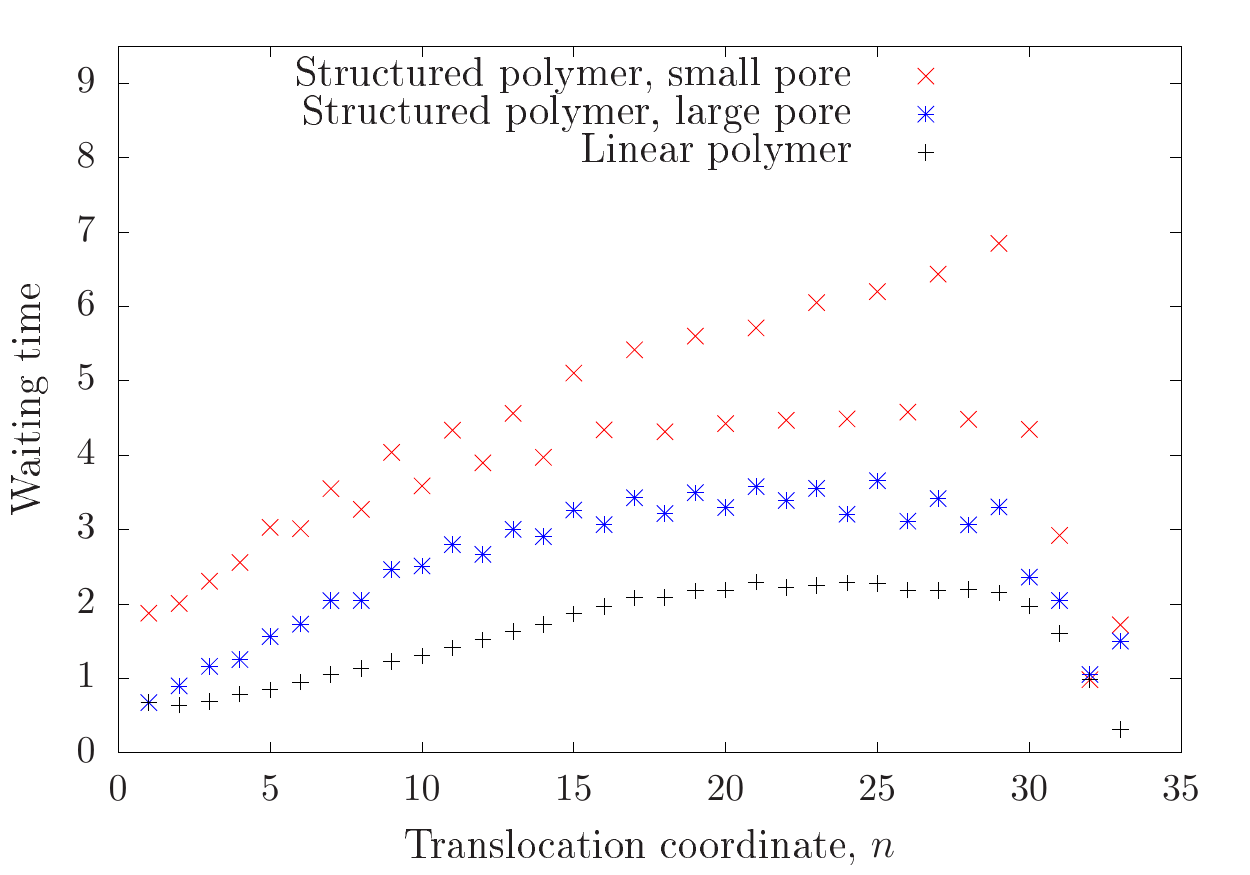}
\caption{Average waiting time depending on the translocation coordinate at given force 
($F=12$) and polymer chain length ($N=32$) for our linear polymer (black) and our 
structured polymer in the case of a large (blue) and small (red) pore. Polymer 
structure interactions with the small pore increases strongly the waiting time 
for odd translocation coordinates. This effect occurs also slightly in the case 
of a large pore.}
\label{FigWaitingtimeStructured}
\end{center}
\end{figure}

The case of the structured polymer translocating through a narrow pore
differs at low forces as we have already mentioned in previous work~\cite{Menais2016}. 
After a look at the mean translocation time (see Figure~\ref{FigComppores}), 
there seems to be almost no difference at high pulling forces.
However when decreasing the pulling force, the mean 
translocation time increases more significantly for a narrow pore 
compared to the previous large pore case and no diffusive regime is observed. 
This increase can reach up to one order of magnitude and may be easily explained 
by an increased interaction (or friction) between the polymer and the pore. 
Those results confirm those already observed for a simple polymer and a narrow pore~\cite{Huopaniemi2007}. Polymer structure interactions with the pore can 
be observed through the waiting times on Figure~\ref{FigWaitingtimeStructured}. 
Although a slight effect of structure is observed for the large pore the case 
of the small pore clearly demonstrates that the translocation process is highly 
slowed down for (odd) grains linked with a lateral grain. This clear effect 
disappears with increasing forces. The shift that can be observed at 
the origin of waiting times for the structured polymer in 
Figure~\ref{FigWaitingtimeStructured} suggest that indeed the term in $O(N)$ 
for the translocation comes from frictions with the pore. 

\begin{figure}[t]
\begin{center}
\includegraphics[width=0.5\textwidth]{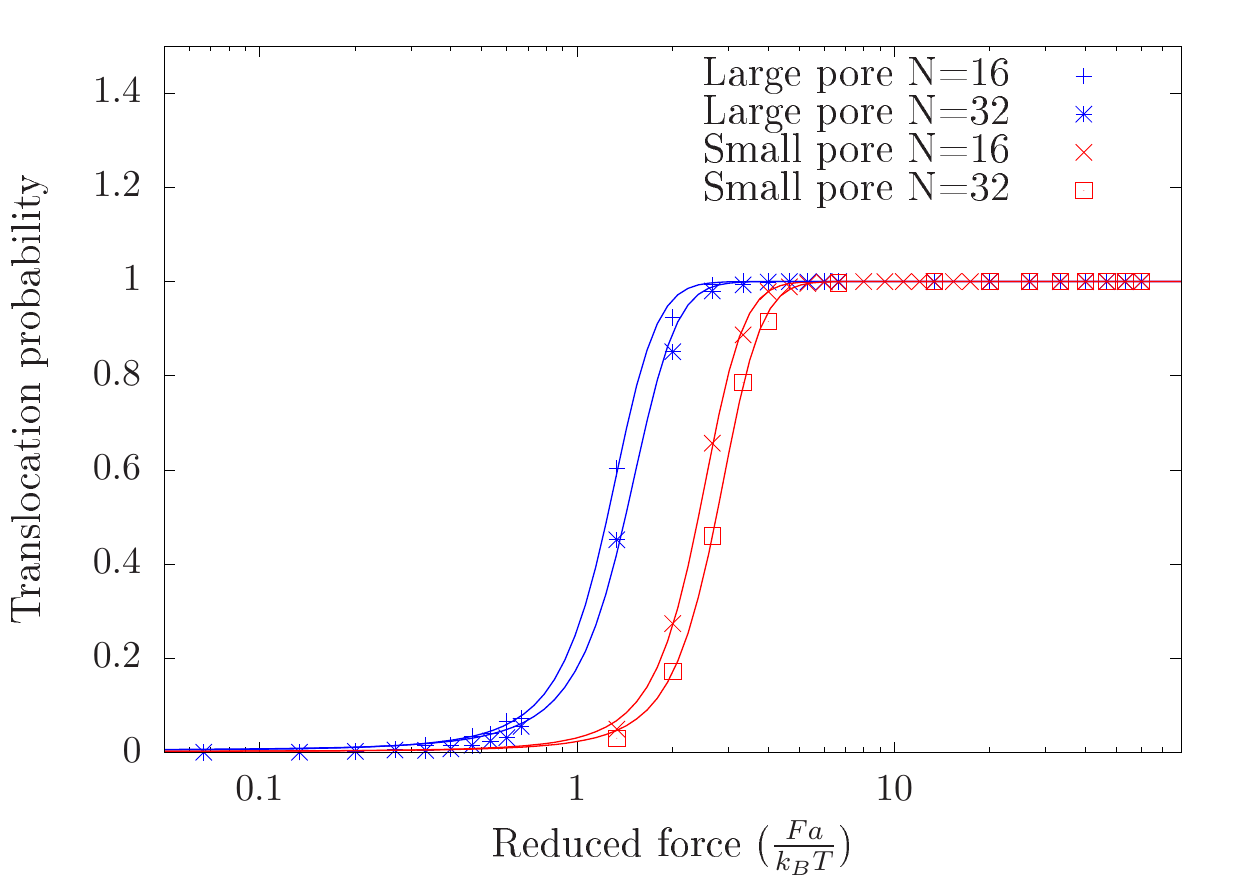}
\caption{Successful translocation probability of the structured polymer in 
function of the reduced force ($Fa/K_BT$). For both pores, the translocation 
is a success at a given pulling force. 
In the case of the narrow pore, one needs to add a threshold reduced force 
(between 1 and 2) to allow the translocation process to take place at a same 
success rate as in the large pore case.}
\label{FigTranslocproba}
\end{center}
\end{figure}

The critical exponent $\delta = 1.72$ observed for the low force regime is also 
strongly different than the usual $\delta \simeq 1$. Our simulations could not 
probe the very low forces as the translocation success probability is too 
small in this regime (see Figure ~\ref{FigTranslocproba}). Instead of a power
law dependence, an exponential decrease of the translocation time as suggested 
in~\cite{Bacci2012} may also be expected as function of the force for the narrow 
pore in case of thermally activated translocation process. In fact, we will show 
in the following that there exists a threshold force below which the translocation 
process is strongly reduced. 
\begin{figure}[t]
\begin{center}
\includegraphics[width=0.5\textwidth]{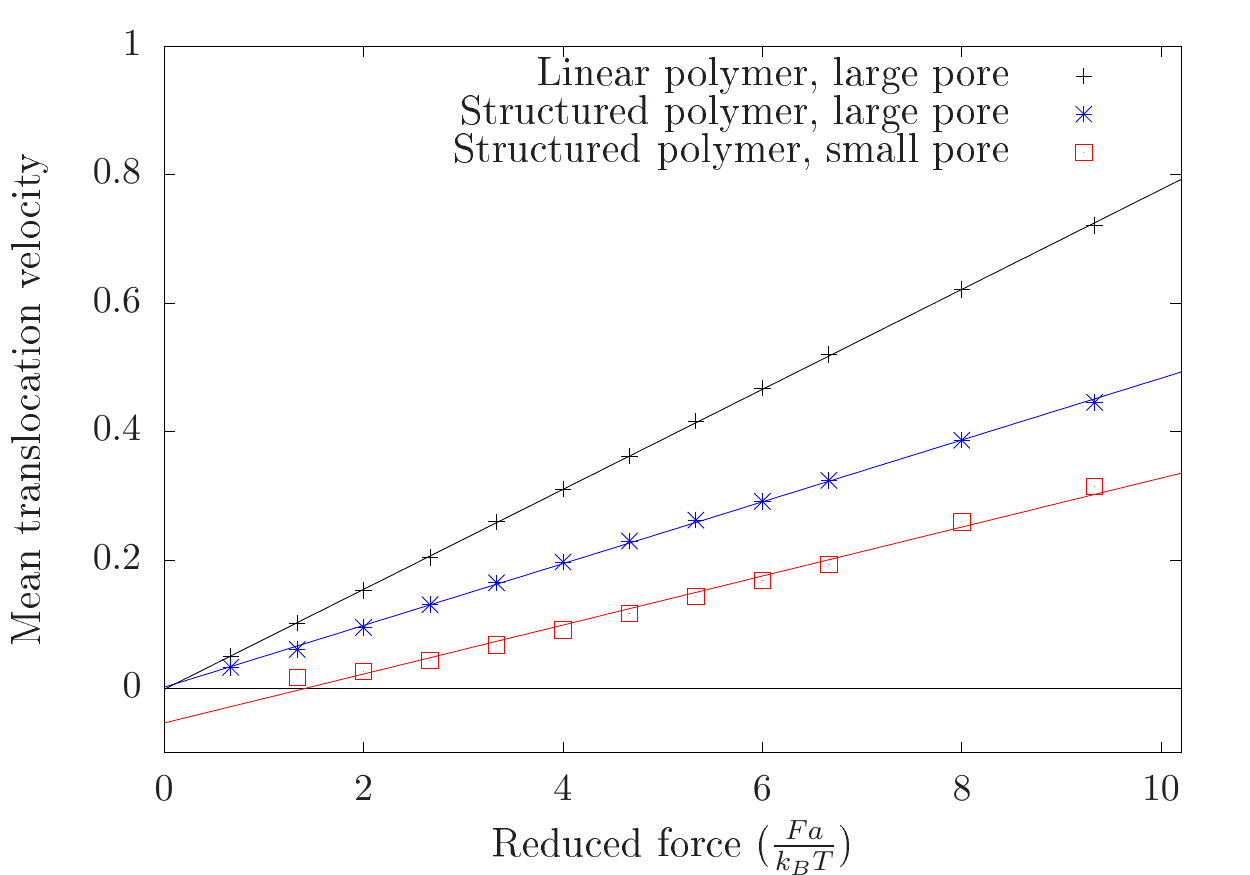}
\caption{Mean translocation velocity for the three studied cases at 
a given chain length ($N=32$) in function of the reduced force ($Fa/K_BT$). 
For large pores we have a linear dependency with a zero ordinate at the origin, 
the factor $3/2$ difference in slope between the linear and structured polymer
originates from the lateral grains friction contribution. For the translocation 
of the structured polymer through a narrow pore, the slope is similar to 
the same polymer through the large pore but with a negative ordinate at the origin.
implying a necessary threshold reduced force to initiate the translocation process.}
\label{FigTranslocvelocities}
\end{center}
\end{figure}

The strong increase of the translocation time for narrow pore may be interpreted by 
a strong friction of the polymer with the narrow pore. This effect is prevalent for 
small forces but almost negligible for high bias forces where the energy brought by 
the force is sufficient to circumvent this friction. In order to test this hypothesis, 
we considered the mean translocation velocity for our three polymer models. 
This velocity is obtained by dividing the mean translocation time by the polymer 
chain length (see Figure~\ref{FigTranslocvelocities}). For large pores, the 
translocation velocity ($v_{lp}$) scales linearly with the pulling force:
\begin{center}
\begin{eqnarray}
 v_{lp} = \xi F
\label{EqFrcitiontransloclarge}
\end{eqnarray}
\end{center}
Unsurprisingly we found that in the case of the structured polymer, 
the translocation velocity is divided by $3/2$ given the supplementary 
lateral grains friction. 
 
In the case of the structured polymer translocation through a narrow pore, 
the mean translocation velocity has the same slope as the same polymer 
through the large pore but shifted by a positive threshold force.
Thus, for the narrow pore mean translocation velocity ($v_{np}$), 
eq.~\ref{EqFrcitiontransloclarge} becomes:
\begin{center}
\begin{eqnarray}
 v_{np} = \xi (F-F^*)
\label{EqFrcitiontranslocnarrow}
\end{eqnarray}
\end{center}
The rescaled threshold force $F^*a/k_B T$ is comprised between 1.5 and 2. 
The same value for the threshold force necessary to start the translocation 
process is observed when we consider the probability of successful 
translocation (defined as the fraction of translocation trial 
ended in the trans side) (see Figure~\ref{FigTranslocproba}).

The friction in the case of a large pore is almost entirely imputable to 
the solvent leading to a viscous friction whereas the contribution of a 
narrow pore is closer to the combination of both a viscous friction (independent 
of the pore size) and a solid-solid friction implying a threshold force necessary 
to translocate the polymer which is pore size dependent.

Structure effects and force thresholds have already been spotted in the literature 
but the mechanism behind those results are very different. The closest to our setup 
(and providing statistical confidence) is a work focused on RNA hairpins~\cite{Vocks2009}, 
the translocation is also biased by pulling but at constant speed instead of constant 
force. The authors determine a threshold force above which the structure is destroyed 
(i.e. the hairpin unzips), thus enabling the translocation to occur. Unfortunately, 
the use of a constant pulling speed prevents the identification of a solid-solid 
friction behavior. Structure and threshold forces were also observed in the case of 
knotted proteins~\cite{Szymczak2016} but in this case, the translocation becomes 
impossible at large pulling forces because the knot is tightened and does 
not slide along the protein anymore. To our knowledge, solid-solid friction behaviour 
induced by polymer structure-pore interactions is an original result in the polymer 
translocation field.

\section{Conclusions}
In this paper we have analyzed the average translocation time of a simple 
linear polymer through a large pore. This reference is then 
used to analyze the effect of a structured polymer on the translocation 
process in particular in case of large and narrow pores. 
We proposed a scaling analysis of characteristic velocities implied in the process
as well as the waiting time as function of the translocation coordinate.
A force driven and a diffusive-like regime were established and the sum over the 
corresponding waiting times of the $N$ monomers leads to the critical exponent
$\alpha = 2$ with finite size effects. This leads us to a confirmation 
of the already found $\alpha=2$ behaviour with finite size effects that may 
explain the underestimation of the effective critical exponent $\alpha$
observed in previous and present numerical simulations. 

The translocation of the structured polymer through a large nanopore showed 
small difference with the linear polymer behaviour except for a slowing down 
of the translocation process by a factor $3/2$ corresponding to the 
supplementary friction coming from the lateral grains. This interesting results
is in agreement with the linearity of all the interactions 
(given that we do not take into account hydrodynamics interactions).

Concerning the translocation of the structured polymer through a narrow 
nanopore, we have shown that at high force regime, the scaling behaviour 
of the mean translocation time is the same compared to a large pore. 
As the force is reduced, no relevant value for $\alpha$ can be established. 
A sharp increase in the value of the critical exponent $\delta$ occurs linked 
to a strong decrease in the successful translocation probability at a 
finite threshold force. This threshold force is also apparent in the linear
behaviour of the mean translocation velocity. The existence of this threshold 
force without modification in the slope of the mean translocation speed exhibits 
a solid-solid friction behavior for structure-pore interactions. 

\section*{Appendix}
Here we develop the calculations including a velocity scale analysis 
using the waiting time function. 

We assume that over the translocation, diffusion is slow 
and the polymer is always quasi-equilibrated in the cis side. Thus, we consider 
separately the trans side fully stretched and the cis side undergoing diffusion. 
This separation is consistant with the polymer conformation on both side as 
shown on Figure~\ref{FigModel}.a. We compare two velocity scales 
corresponding to the cis and trans sides. Concerning the cis side, 
the diffusion velocity originates from the Brownian motion through energy 
equipartition of the $N-n$ monomers present in the cis side: 
\begin{equation}
v_{cis} \propto \frac{\sqrt{\nu(N-n)k_BT}}{\nu(N-n)} = {\sqrt{\frac{k_BT}{\nu(N-n)}}}.
\end{equation}
For the trans side, two velocity scales exist: the same Brownian velocity 
which is relevant only for very small bias forces, i.e. in the quasi-unbiased
regime (see Figure~\ref{FigLinpolTranslocationtime}) and the velocity 
originating from the pulling force: 
\begin{equation}
v_{trans} \propto \frac{F}{\nu n}.
\end{equation}
The transition between the two first regimes observed for the waiting time 
occurs when both velocities equalize $v_{trans} \approx v_{cis}$. 
The crossover between the first regime where the transolcation is imposed 
by the pulling force and the second regime of quasi-unbiased translocation 
occurs for $n$ monomers translocated at a force $F(n)$: 
\begin{equation}
F(n) \propto \sqrt{\frac{\nu k_B T n^2}{N-n}}.
\label{eqforce}
\end{equation}

Equation~\ref{eqforce} may be simply inverted to get the monomer 
$n^*(F,N)$ at which 
the crossover between the force driven and diffusive regime occurs 
for a given pulling force $F$ and polymer length $N$:
\begin{equation}
n^*(F,N)=\frac{F^2}{2C}\sqrt {1+\frac{4NC}{F^2}}-\frac{F^2}{2C}
\label{EqBoundary}
\end{equation}
where $C$ is a constant. For large pulling forces, when $F \gg \sqrt{CN}$, 
$n^* \rightarrow N$, meaning that the force driven regime always predominates.
For low forces, $n^* \rightarrow 1$ and the unbiased translocation regime is 
recovered on the whole translocation.
The fraction of the polymer corresponding to the crossover, $\beta(F,N)=n^*/N$ 
is only function of the reduced parameter $A=2CN/F^2$:
\begin{equation}
\beta(F,N) = \beta(A)= \frac{\sqrt{1+2A}-1}{A}.
\end{equation}
As previously stated, the mean translocation time is the sum over all 
the translocation coordinates of the average waiting times $w(n)$. 
It may be reduced to the integration over the two separate regimes 
if one neglects the tail retraction.
From Figure~\ref{FigLinpolwaitingtime}, the waiting time $w(n)$
is linear for the force driven regime with a constant term $a(F)$ 
and a slope $b(F)$. The second diffusive regime corresponds to a 
quasi-constant waiting time although slightly decreasing when $(N-n) \gg 1$ 
and sharply decreasing on 4-5 monomers for the third regime of polymer tail 
retraction. Thus, we obtain equation \ref{EqIntegral} mentioned in the main text.
Numerical evidence show that $a(F)$ and $b(F)$ are inversely 
proportional to $F$\footnotemark[1], hence:
\begin{center}
\begin{eqnarray}
\tau(F,N) = \frac{a'}{F} N + \left(\beta(A)-\frac{\beta(A)^2}{2}\right) 
\frac{b'N^2}{F}
\label{EqTauformulaappendix}
\end{eqnarray}
\end{center}
with $a'$ and $b'$ constants. For a reasonable couple of $N$ and $F$ values 
($A < 1$), the pre-factor in front of the term scaling as $N^2/F$ remains 
a constant as shown in SM \footnotemark[1]. We then have equation \ref{EqTauformula} valid provided the applied force is large enough to avoid highly diffusive regimes (i.e. $1\ll \sqrt{2CN} < F$). The expected scaling  goes from $\tau \propto N^{2+\nu}$ for the unbiased case to $\tau \propto N^{2}/F$ in the force driven case far from finite size effects.

\section*{Acknowledgments}
We thank L\'eo Brunswick for fruitful discussions. Grand Équipement National de Calcul Intensif (GENCI) for the allocation of computing resources INAC-SYMMES is part of the Arcane Labex program, funded by the French National Research Agency (ARCANE project no. ANR-12-LABX-003).

\bibliographystyle{apsrev4-1}
\bibliography{articlebib}

\begin{thebibliography}{56}%
\makeatletter
\providecommand \@ifxundefined [1]{%
 \@ifx{#1\undefined}
}%
\providecommand \@ifnum [1]{%
 \ifnum #1\expandafter \@firstoftwo
 \else \expandafter \@secondoftwo
 \fi
}%
\providecommand \@ifx [1]{%
 \ifx #1\expandafter \@firstoftwo
 \else \expandafter \@secondoftwo
 \fi
}%
\providecommand \natexlab [1]{#1}%
\providecommand \enquote  [1]{``#1''}%
\providecommand \bibnamefont  [1]{#1}%
\providecommand \bibfnamefont [1]{#1}%
\providecommand \citenamefont [1]{#1}%
\providecommand \href@noop [0]{\@secondoftwo}%
\providecommand \href [0]{\begingroup \@sanitize@url \@href}%
\providecommand \@href[1]{\@@startlink{#1}\@@href}%
\providecommand \@@href[1]{\endgroup#1\@@endlink}%
\providecommand \@sanitize@url [0]{\catcode `\\12\catcode `\$12\catcode
  `\&12\catcode `\#12\catcode `\^12\catcode `\_12\catcode `\%12\relax}%
\providecommand \@@startlink[1]{}%
\providecommand \@@endlink[0]{}%
\providecommand \url  [0]{\begingroup\@sanitize@url \@url }%
\providecommand \@url [1]{\endgroup\@href {#1}{\urlprefix }}%
\providecommand \urlprefix  [0]{URL }%
\providecommand \Eprint [0]{\href }%
\providecommand \doibase [0]{http://dx.doi.org/}%
\providecommand \selectlanguage [0]{\@gobble}%
\providecommand \bibinfo  [0]{\@secondoftwo}%
\providecommand \bibfield  [0]{\@secondoftwo}%
\providecommand \translation [1]{[#1]}%
\providecommand \BibitemOpen [0]{}%
\providecommand \bibitemStop [0]{}%
\providecommand \bibitemNoStop [0]{.\EOS\space}%
\providecommand \EOS [0]{\spacefactor3000\relax}%
\providecommand \BibitemShut  [1]{\csname bibitem#1\endcsname}%
\let\auto@bib@innerbib\@empty
\bibitem [{\citenamefont {Kasianowicz}\ \emph {et~al.}(1996)\citenamefont
  {Kasianowicz}, \citenamefont {Brandin}, \citenamefont {Branton},\ and\
  \citenamefont {Deamer}}]{Kasianowicz1996}%
  \BibitemOpen
  \bibfield  {author} {\bibinfo {author} {\bibfnamefont {J.~J.}\ \bibnamefont
  {Kasianowicz}}, \bibinfo {author} {\bibfnamefont {E.}~\bibnamefont
  {Brandin}}, \bibinfo {author} {\bibfnamefont {D.}~\bibnamefont {Branton}}, \
  and\ \bibinfo {author} {\bibfnamefont {D.~W.}\ \bibnamefont {Deamer}},\
  }\href {\doibase 10.1073/pnas.93.24.13770} {\bibfield  {journal} {\bibinfo
  {journal} {Proc. Natl. Acad. Sc. USA}\ }\textbf {\bibinfo {volume} {93}},\
  \bibinfo {pages} {13770} (\bibinfo {year} {1996})}\BibitemShut {NoStop}%
\bibitem [{\citenamefont {Dekker}(2007)}]{Dekker2007}%
  \BibitemOpen
  \bibfield  {author} {\bibinfo {author} {\bibfnamefont {C.}~\bibnamefont
  {Dekker}},\ }\href {\doibase 10.1038/nnano.2007.27} {\bibfield  {journal}
  {\bibinfo  {journal} {Nature Nanotech}\ }\textbf {\bibinfo {volume} {2}},\
  \bibinfo {pages} {209} (\bibinfo {year} {2007})}\BibitemShut {NoStop}%
\bibitem [{\citenamefont {Keyser}(2011)}]{Keyser}%
  \BibitemOpen
  \bibfield  {author} {\bibinfo {author} {\bibfnamefont {U.~F.}\ \bibnamefont
  {Keyser}},\ }\href {\doibase 10.1098/rsif.2011.0222} {\bibfield  {journal}
  {\bibinfo  {journal} {J. Royal Soc. Interface}\ }\textbf {\bibinfo {volume}
  {8}},\ \bibinfo {pages} {1369} (\bibinfo {year} {2011})}\BibitemShut
  {NoStop}%
\bibitem [{\citenamefont {Meller}(2003)}]{Meller2003}%
  \BibitemOpen
  \bibfield  {author} {\bibinfo {author} {\bibfnamefont {A.}~\bibnamefont
  {Meller}},\ }\href {\doibase 10.1088/0953-8984/15/17/202} {\bibfield
  {journal} {\bibinfo  {journal} {J. Phys.: Condensed Matter}\ }\textbf
  {\bibinfo {volume} {15}},\ \bibinfo {pages} {R581} (\bibinfo {year}
  {2003})}\BibitemShut {NoStop}%
\bibitem [{\citenamefont {Movileanu}(2008)}]{Movileanu2008}%
  \BibitemOpen
  \bibfield  {author} {\bibinfo {author} {\bibfnamefont {L.}~\bibnamefont
  {Movileanu}},\ }\href {\doibase 10.1039/b719850g} {\bibfield  {journal}
  {\bibinfo  {journal} {Soft Matter}\ }\textbf {\bibinfo {volume} {4}},\
  \bibinfo {pages} {925} (\bibinfo {year} {2008})}\BibitemShut {NoStop}%
\bibitem [{\citenamefont {Palyulin}\ \emph {et~al.}(2014)\citenamefont
  {Palyulin}, \citenamefont {Ala-Nissila},\ and\ \citenamefont
  {Metzler}}]{Palyulin2014}%
  \BibitemOpen
  \bibfield  {author} {\bibinfo {author} {\bibfnamefont {V.~V.}\ \bibnamefont
  {Palyulin}}, \bibinfo {author} {\bibfnamefont {T.}~\bibnamefont
  {Ala-Nissila}}, \ and\ \bibinfo {author} {\bibfnamefont {R.}~\bibnamefont
  {Metzler}},\ }\href {\doibase 10.1039/c4sm01819b} {\bibfield  {journal}
  {\bibinfo  {journal} {Soft Matter}\ }\textbf {\bibinfo {volume} {10}},\
  \bibinfo {pages} {9016} (\bibinfo {year} {2014})}\BibitemShut {NoStop}%
\bibitem [{\citenamefont {Wanunu}(2012)}]{Wanunu2012}%
  \BibitemOpen
  \bibfield  {author} {\bibinfo {author} {\bibfnamefont {M.}~\bibnamefont
  {Wanunu}},\ }\href {\doibase 10.1016/j.plrev.2012.05.010} {\bibfield
  {journal} {\bibinfo  {journal} {Physics of Life Reviews}\ }\textbf {\bibinfo
  {volume} {9}},\ \bibinfo {pages} {125} (\bibinfo {year} {2012})}\BibitemShut
  {NoStop}%
\bibitem [{\citenamefont {Liu}\ \emph {et~al.}(2013)\citenamefont {Liu},
  \citenamefont {Bombera}, \citenamefont {Leroy}, \citenamefont {Roupioz},
  \citenamefont {Baganizi}, \citenamefont {Marche}, \citenamefont {Haguet},
  \citenamefont {Mailley},\ and\ \citenamefont {Livache}}]{Liu2013}%
  \BibitemOpen
  \bibfield  {author} {\bibinfo {author} {\bibfnamefont {J.}~\bibnamefont
  {Liu}}, \bibinfo {author} {\bibfnamefont {R.}~\bibnamefont {Bombera}},
  \bibinfo {author} {\bibfnamefont {L.}~\bibnamefont {Leroy}}, \bibinfo
  {author} {\bibfnamefont {Y.}~\bibnamefont {Roupioz}}, \bibinfo {author}
  {\bibfnamefont {D.~R.}\ \bibnamefont {Baganizi}}, \bibinfo {author}
  {\bibfnamefont {P.~N.}\ \bibnamefont {Marche}}, \bibinfo {author}
  {\bibfnamefont {V.}~\bibnamefont {Haguet}}, \bibinfo {author} {\bibfnamefont
  {P.}~\bibnamefont {Mailley}}, \ and\ \bibinfo {author} {\bibfnamefont
  {T.}~\bibnamefont {Livache}},\ }\href {\doibase 10.1371/journal.pone.0057717}
  {\bibfield  {journal} {\bibinfo  {journal} {{PLoS} {ONE}}\ }\textbf {\bibinfo
  {volume} {8}},\ \bibinfo {pages} {e57717} (\bibinfo {year}
  {2013})}\BibitemShut {NoStop}%
\bibitem [{\citenamefont {Branton}\ \emph {et~al.}(2008)\citenamefont
  {Branton}, \citenamefont {Deamer}, \citenamefont {Marziali}, \citenamefont
  {Bayley}, \citenamefont {Benner}, \citenamefont {Butler}, \citenamefont
  {Ventra}, \citenamefont {Garaj}, \citenamefont {Hibbs}, \citenamefont
  {Huang}, \citenamefont {Jovanovich}, \citenamefont {Krstic}, \citenamefont
  {Lindsay}, \citenamefont {Ling}, \citenamefont {Mastrangelo}, \citenamefont
  {Meller}, \citenamefont {Oliver}, \citenamefont {Pershin}, \citenamefont
  {Ramsey}, \citenamefont {Riehn}, \citenamefont {Soni}, \citenamefont
  {Tabard-Cossa}, \citenamefont {Wanunu}, \citenamefont {Wiggin},\ and\
  \citenamefont {Schloss}}]{Branton2008}%
  \BibitemOpen
  \bibfield  {author} {\bibinfo {author} {\bibfnamefont {D.}~\bibnamefont
  {Branton}}, \bibinfo {author} {\bibfnamefont {D.~W.}\ \bibnamefont {Deamer}},
  \bibinfo {author} {\bibfnamefont {A.}~\bibnamefont {Marziali}}, \bibinfo
  {author} {\bibfnamefont {H.}~\bibnamefont {Bayley}}, \bibinfo {author}
  {\bibfnamefont {S.~A.}\ \bibnamefont {Benner}}, \bibinfo {author}
  {\bibfnamefont {T.}~\bibnamefont {Butler}}, \bibinfo {author} {\bibfnamefont
  {M.~D.}\ \bibnamefont {Ventra}}, \bibinfo {author} {\bibfnamefont
  {S.}~\bibnamefont {Garaj}}, \bibinfo {author} {\bibfnamefont
  {A.}~\bibnamefont {Hibbs}}, \bibinfo {author} {\bibfnamefont
  {X.}~\bibnamefont {Huang}}, \bibinfo {author} {\bibfnamefont {S.~B.}\
  \bibnamefont {Jovanovich}}, \bibinfo {author} {\bibfnamefont {P.~S.}\
  \bibnamefont {Krstic}}, \bibinfo {author} {\bibfnamefont {S.}~\bibnamefont
  {Lindsay}}, \bibinfo {author} {\bibfnamefont {X.~S.}\ \bibnamefont {Ling}},
  \bibinfo {author} {\bibfnamefont {C.~H.}\ \bibnamefont {Mastrangelo}},
  \bibinfo {author} {\bibfnamefont {A.}~\bibnamefont {Meller}}, \bibinfo
  {author} {\bibfnamefont {J.~S.}\ \bibnamefont {Oliver}}, \bibinfo {author}
  {\bibfnamefont {Y.~V.}\ \bibnamefont {Pershin}}, \bibinfo {author}
  {\bibfnamefont {J.~M.}\ \bibnamefont {Ramsey}}, \bibinfo {author}
  {\bibfnamefont {R.}~\bibnamefont {Riehn}}, \bibinfo {author} {\bibfnamefont
  {G.~V.}\ \bibnamefont {Soni}}, \bibinfo {author} {\bibfnamefont
  {V.}~\bibnamefont {Tabard-Cossa}}, \bibinfo {author} {\bibfnamefont
  {M.}~\bibnamefont {Wanunu}}, \bibinfo {author} {\bibfnamefont
  {M.}~\bibnamefont {Wiggin}}, \ and\ \bibinfo {author} {\bibfnamefont {J.~A.}\
  \bibnamefont {Schloss}},\ }\href {\doibase 10.1038/nbt.1495} {\bibfield
  {journal} {\bibinfo  {journal} {Nat Biotechnol}\ }\textbf {\bibinfo {volume}
  {26}},\ \bibinfo {pages} {1146} (\bibinfo {year} {2008})}\BibitemShut
  {NoStop}%
\bibitem [{\citenamefont {Clarke}\ \emph {et~al.}(2009)\citenamefont {Clarke},
  \citenamefont {Wu}, \citenamefont {Jayasinghe}, \citenamefont {Patel},
  \citenamefont {Reid},\ and\ \citenamefont {Bayley}}]{Clarke2009}%
  \BibitemOpen
  \bibfield  {author} {\bibinfo {author} {\bibfnamefont {J.}~\bibnamefont
  {Clarke}}, \bibinfo {author} {\bibfnamefont {H.-C.}\ \bibnamefont {Wu}},
  \bibinfo {author} {\bibfnamefont {L.}~\bibnamefont {Jayasinghe}}, \bibinfo
  {author} {\bibfnamefont {A.}~\bibnamefont {Patel}}, \bibinfo {author}
  {\bibfnamefont {S.}~\bibnamefont {Reid}}, \ and\ \bibinfo {author}
  {\bibfnamefont {H.}~\bibnamefont {Bayley}},\ }\href {\doibase
  10.1038/nnano.2009.12} {\bibfield  {journal} {\bibinfo  {journal} {Nature
  Nanotech}\ }\textbf {\bibinfo {volume} {4}},\ \bibinfo {pages} {265}
  (\bibinfo {year} {2009})}\BibitemShut {NoStop}%
\bibitem [{\citenamefont {Merchant}\ \emph {et~al.}(2010)\citenamefont
  {Merchant}, \citenamefont {Healy}, \citenamefont {Wanunu}, \citenamefont
  {Ray}, \citenamefont {Peterman}, \citenamefont {Bartel}, \citenamefont
  {Fischbein}, \citenamefont {Venta}, \citenamefont {Luo}, \citenamefont
  {Johnson},\ and\ \citenamefont {Drndi{\'c}}}]{Merchant2010}%
  \BibitemOpen
  \bibfield  {author} {\bibinfo {author} {\bibfnamefont {C.~A.}\ \bibnamefont
  {Merchant}}, \bibinfo {author} {\bibfnamefont {K.}~\bibnamefont {Healy}},
  \bibinfo {author} {\bibfnamefont {M.}~\bibnamefont {Wanunu}}, \bibinfo
  {author} {\bibfnamefont {V.}~\bibnamefont {Ray}}, \bibinfo {author}
  {\bibfnamefont {N.}~\bibnamefont {Peterman}}, \bibinfo {author}
  {\bibfnamefont {J.}~\bibnamefont {Bartel}}, \bibinfo {author} {\bibfnamefont
  {M.~D.}\ \bibnamefont {Fischbein}}, \bibinfo {author} {\bibfnamefont
  {K.}~\bibnamefont {Venta}}, \bibinfo {author} {\bibfnamefont
  {Z.}~\bibnamefont {Luo}}, \bibinfo {author} {\bibfnamefont {A.~T.~C.}\
  \bibnamefont {Johnson}}, \ and\ \bibinfo {author} {\bibfnamefont
  {M.}~\bibnamefont {Drndi{\'c}}},\ }\href {\doibase 10.1021/nl101046t}
  {\bibfield  {journal} {\bibinfo  {journal} {Nano Letters}\ }\textbf {\bibinfo
  {volume} {10}},\ \bibinfo {pages} {2915} (\bibinfo {year}
  {2010})}\BibitemShut {NoStop}%
\bibitem [{\citenamefont {Schneider}\ and\ \citenamefont
  {Dekker}(2012)}]{Schneider2012}%
  \BibitemOpen
  \bibfield  {author} {\bibinfo {author} {\bibfnamefont {G.~F.}\ \bibnamefont
  {Schneider}}\ and\ \bibinfo {author} {\bibfnamefont {C.}~\bibnamefont
  {Dekker}},\ }\href {\doibase 10.1038/nbt.2181} {\bibfield  {journal}
  {\bibinfo  {journal} {Nat Biotechnol}\ }\textbf {\bibinfo {volume} {30}},\
  \bibinfo {pages} {326} (\bibinfo {year} {2012})}\BibitemShut {NoStop}%
\bibitem [{\citenamefont {Schneider}\ \emph {et~al.}(2010)\citenamefont
  {Schneider}, \citenamefont {Kowalczyk}, \citenamefont {Calado}, \citenamefont
  {Pandraud}, \citenamefont {Zandbergen}, \citenamefont {Vandersypen},\ and\
  \citenamefont {Dekker}}]{Schneider2010}%
  \BibitemOpen
  \bibfield  {author} {\bibinfo {author} {\bibfnamefont {G.~F.}\ \bibnamefont
  {Schneider}}, \bibinfo {author} {\bibfnamefont {S.~W.}\ \bibnamefont
  {Kowalczyk}}, \bibinfo {author} {\bibfnamefont {V.~E.}\ \bibnamefont
  {Calado}}, \bibinfo {author} {\bibfnamefont {G.}~\bibnamefont {Pandraud}},
  \bibinfo {author} {\bibfnamefont {H.~W.}\ \bibnamefont {Zandbergen}},
  \bibinfo {author} {\bibfnamefont {L.~M.~K.}\ \bibnamefont {Vandersypen}}, \
  and\ \bibinfo {author} {\bibfnamefont {C.}~\bibnamefont {Dekker}},\ }\href
  {\doibase 10.1021/nl102069z} {\bibfield  {journal} {\bibinfo  {journal} {Nano
  Letters}\ }\textbf {\bibinfo {volume} {10}},\ \bibinfo {pages} {3163}
  (\bibinfo {year} {2010})}\BibitemShut {NoStop}%
\bibitem [{\citenamefont {Kantor}\ and\ \citenamefont
  {Kardar}(2004)}]{Kantor2004}%
  \BibitemOpen
  \bibfield  {author} {\bibinfo {author} {\bibfnamefont {Y.}~\bibnamefont
  {Kantor}}\ and\ \bibinfo {author} {\bibfnamefont {M.}~\bibnamefont
  {Kardar}},\ }\href {\doibase 10.1103/physreve.69.021806} {\bibfield
  {journal} {\bibinfo  {journal} {Phys. Rev. E}\ }\textbf {\bibinfo {volume}
  {69}},\ \bibinfo {pages} {021806} (\bibinfo {year} {2004})}\BibitemShut
  {NoStop}%
\bibitem [{\citenamefont {Keyser}\ \emph {et~al.}(2006)\citenamefont {Keyser},
  \citenamefont {Koeleman}, \citenamefont {Dorp}, \citenamefont {Krapf},
  \citenamefont {Smeets}, \citenamefont {Lemay}, \citenamefont {Dekker},\ and\
  \citenamefont {Dekker}}]{Keyser2006}%
  \BibitemOpen
  \bibfield  {author} {\bibinfo {author} {\bibfnamefont {U.~F.}\ \bibnamefont
  {Keyser}}, \bibinfo {author} {\bibfnamefont {B.~N.}\ \bibnamefont
  {Koeleman}}, \bibinfo {author} {\bibfnamefont {S.~V.}\ \bibnamefont {Dorp}},
  \bibinfo {author} {\bibfnamefont {D.}~\bibnamefont {Krapf}}, \bibinfo
  {author} {\bibfnamefont {R.~M.~M.}\ \bibnamefont {Smeets}}, \bibinfo {author}
  {\bibfnamefont {S.~G.}\ \bibnamefont {Lemay}}, \bibinfo {author}
  {\bibfnamefont {N.~H.}\ \bibnamefont {Dekker}}, \ and\ \bibinfo {author}
  {\bibfnamefont {C.}~\bibnamefont {Dekker}},\ }\href@noop {} {\bibfield
  {journal} {\bibinfo  {journal} {Nature Physics}\ }\textbf {\bibinfo {volume}
  {2}},\ \bibinfo {pages} {473} (\bibinfo {year} {2006})}\BibitemShut {NoStop}%
\bibitem [{\citenamefont {Peng}\ and\ \citenamefont {Ling}(2009)}]{Peng2009}%
  \BibitemOpen
  \bibfield  {author} {\bibinfo {author} {\bibfnamefont {H.}~\bibnamefont
  {Peng}}\ and\ \bibinfo {author} {\bibfnamefont {X.~S.}\ \bibnamefont
  {Ling}},\ }\href {\doibase 10.1088/0957-4484/20/18/185101} {\bibfield
  {journal} {\bibinfo  {journal} {Nanotechnology}\ }\textbf {\bibinfo {volume}
  {20}},\ \bibinfo {pages} {185101} (\bibinfo {year} {2009})}\BibitemShut
  {NoStop}%
\bibitem [{\citenamefont {van Dorp}\ \emph {et~al.}(2009)\citenamefont {van
  Dorp}, \citenamefont {Keyser}, \citenamefont {Dekker}, \citenamefont
  {Dekker},\ and\ \citenamefont {Lemay}}]{VanDorp2009}%
  \BibitemOpen
  \bibfield  {author} {\bibinfo {author} {\bibfnamefont {S.}~\bibnamefont {van
  Dorp}}, \bibinfo {author} {\bibfnamefont {U.~F.}\ \bibnamefont {Keyser}},
  \bibinfo {author} {\bibfnamefont {N.~H.}\ \bibnamefont {Dekker}}, \bibinfo
  {author} {\bibfnamefont {C.}~\bibnamefont {Dekker}}, \ and\ \bibinfo {author}
  {\bibfnamefont {S.~G.}\ \bibnamefont {Lemay}},\ }\href {\doibase
  10.1038/nphys1230} {\bibfield  {journal} {\bibinfo  {journal} {Nat Phys}\
  }\textbf {\bibinfo {volume} {5}},\ \bibinfo {pages} {347} (\bibinfo {year}
  {2009})}\BibitemShut {NoStop}%
\bibitem [{\citenamefont {Sischka}\ \emph {et~al.}(2010)\citenamefont
  {Sischka}, \citenamefont {Spiering}, \citenamefont {Khaksar}, \citenamefont
  {Laxa}, \citenamefont {K\"{o}nig}, \citenamefont {Dietz},\ and\ \citenamefont
  {Anselmetti}}]{Sischka2010}%
  \BibitemOpen
  \bibfield  {author} {\bibinfo {author} {\bibfnamefont {A.}~\bibnamefont
  {Sischka}}, \bibinfo {author} {\bibfnamefont {A.}~\bibnamefont {Spiering}},
  \bibinfo {author} {\bibfnamefont {M.}~\bibnamefont {Khaksar}}, \bibinfo
  {author} {\bibfnamefont {M.}~\bibnamefont {Laxa}}, \bibinfo {author}
  {\bibfnamefont {J.}~\bibnamefont {K\"{o}nig}}, \bibinfo {author}
  {\bibfnamefont {K.-J.}\ \bibnamefont {Dietz}}, \ and\ \bibinfo {author}
  {\bibfnamefont {D.}~\bibnamefont {Anselmetti}},\ }\href {\doibase
  10.1088/0953-8984/22/45/454121} {\bibfield  {journal} {\bibinfo  {journal}
  {Journal of Physics: Condensed Matter}\ }\textbf {\bibinfo {volume} {22}},\
  \bibinfo {pages} {454121} (\bibinfo {year} {2010})}\BibitemShut {NoStop}%
\bibitem [{\citenamefont {Bulushev}\ \emph {et~al.}(2014)\citenamefont
  {Bulushev}, \citenamefont {Steinbock}, \citenamefont {Khlybov}, \citenamefont
  {Steinbock}, \citenamefont {Keyser},\ and\ \citenamefont
  {Radenovic}}]{Bulushev2014}%
  \BibitemOpen
  \bibfield  {author} {\bibinfo {author} {\bibfnamefont {R.~D.}\ \bibnamefont
  {Bulushev}}, \bibinfo {author} {\bibfnamefont {L.~J.}\ \bibnamefont
  {Steinbock}}, \bibinfo {author} {\bibfnamefont {S.}~\bibnamefont {Khlybov}},
  \bibinfo {author} {\bibfnamefont {J.~F.}\ \bibnamefont {Steinbock}}, \bibinfo
  {author} {\bibfnamefont {U.~F.}\ \bibnamefont {Keyser}}, \ and\ \bibinfo
  {author} {\bibfnamefont {A.}~\bibnamefont {Radenovic}},\ }\href {\doibase
  10.1021/nl503272r} {\bibfield  {journal} {\bibinfo  {journal} {Nano Letters}\
  }\textbf {\bibinfo {volume} {14}},\ \bibinfo {pages} {6606} (\bibinfo {year}
  {2014})}\BibitemShut {NoStop}%
\bibitem [{\citenamefont {Bulushev}\ \emph {et~al.}(2015)\citenamefont
  {Bulushev}, \citenamefont {Marion},\ and\ \citenamefont
  {Radenovic}}]{Bulushev2015}%
  \BibitemOpen
  \bibfield  {author} {\bibinfo {author} {\bibfnamefont {R.~D.}\ \bibnamefont
  {Bulushev}}, \bibinfo {author} {\bibfnamefont {S.}~\bibnamefont {Marion}}, \
  and\ \bibinfo {author} {\bibfnamefont {A.}~\bibnamefont {Radenovic}},\ }\href
  {\doibase 10.1021/acs.nanolett.5b03264} {\bibfield  {journal} {\bibinfo
  {journal} {Nano Letters}\ }\textbf {\bibinfo {volume} {15}},\ \bibinfo
  {pages} {7118} (\bibinfo {year} {2015})}\BibitemShut {NoStop}%
\bibitem [{\citenamefont {Bulushev}\ \emph {et~al.}(2016)\citenamefont
  {Bulushev}, \citenamefont {Marion}, \citenamefont {Petrova}, \citenamefont
  {Davis}, \citenamefont {Maerkl},\ and\ \citenamefont
  {Radenovic}}]{Bulushev2016}%
  \BibitemOpen
  \bibfield  {author} {\bibinfo {author} {\bibfnamefont {R.~D.}\ \bibnamefont
  {Bulushev}}, \bibinfo {author} {\bibfnamefont {S.}~\bibnamefont {Marion}},
  \bibinfo {author} {\bibfnamefont {E.}~\bibnamefont {Petrova}}, \bibinfo
  {author} {\bibfnamefont {S.~J.}\ \bibnamefont {Davis}}, \bibinfo {author}
  {\bibfnamefont {S.~J.}\ \bibnamefont {Maerkl}}, \ and\ \bibinfo {author}
  {\bibfnamefont {A.}~\bibnamefont {Radenovic}},\ }\href {\doibase
  10.1021/acs.nanolett.6b04165} {\bibfield  {journal} {\bibinfo  {journal}
  {Nano Letters}\ }\textbf {\bibinfo {volume} {16}},\ \bibinfo {pages} {7882}
  (\bibinfo {year} {2016})}\BibitemShut {NoStop}%
\bibitem [{\citenamefont {Yameen}\ \emph {et~al.}(2009)\citenamefont {Yameen},
  \citenamefont {Ali}, \citenamefont {Neumann}, \citenamefont {Ensinger},
  \citenamefont {Knoll},\ and\ \citenamefont {Azzaroni}}]{Yameen2009}%
  \BibitemOpen
  \bibfield  {author} {\bibinfo {author} {\bibfnamefont {B.}~\bibnamefont
  {Yameen}}, \bibinfo {author} {\bibfnamefont {M.}~\bibnamefont {Ali}},
  \bibinfo {author} {\bibfnamefont {R.}~\bibnamefont {Neumann}}, \bibinfo
  {author} {\bibfnamefont {W.}~\bibnamefont {Ensinger}}, \bibinfo {author}
  {\bibfnamefont {W.}~\bibnamefont {Knoll}}, \ and\ \bibinfo {author}
  {\bibfnamefont {O.}~\bibnamefont {Azzaroni}},\ }\href {\doibase
  10.1002/smll.200801318} {\bibfield  {journal} {\bibinfo  {journal} {Small}\
  }\textbf {\bibinfo {volume} {5}},\ \bibinfo {pages} {1287} (\bibinfo {year}
  {2009})}\BibitemShut {NoStop}%
\bibitem [{\citenamefont {Grosberg}\ \emph {et~al.}(2006)\citenamefont
  {Grosberg}, \citenamefont {Nechaev}, \citenamefont {Tamm},\ and\
  \citenamefont {Vasilyev}}]{Grosberg2006}%
  \BibitemOpen
  \bibfield  {author} {\bibinfo {author} {\bibfnamefont {A.~Y.}\ \bibnamefont
  {Grosberg}}, \bibinfo {author} {\bibfnamefont {S.}~\bibnamefont {Nechaev}},
  \bibinfo {author} {\bibfnamefont {M.}~\bibnamefont {Tamm}}, \ and\ \bibinfo
  {author} {\bibfnamefont {O.}~\bibnamefont {Vasilyev}},\ }\href {\doibase
  10.1103/physrevlett.96.228105} {\bibfield  {journal} {\bibinfo  {journal}
  {Phys. Rev. Lett.}\ }\textbf {\bibinfo {volume} {96}},\ \bibinfo {pages}
  {228105} (\bibinfo {year} {2006})}\BibitemShut {NoStop}%
\bibitem [{\citenamefont {Huopaniemi}\ \emph {et~al.}(2007)\citenamefont
  {Huopaniemi}, \citenamefont {Luo}, \citenamefont {Ala-Nissila},\ and\
  \citenamefont {Ying}}]{Huopaniemi2007}%
  \BibitemOpen
  \bibfield  {author} {\bibinfo {author} {\bibfnamefont {I.}~\bibnamefont
  {Huopaniemi}}, \bibinfo {author} {\bibfnamefont {K.}~\bibnamefont {Luo}},
  \bibinfo {author} {\bibfnamefont {T.}~\bibnamefont {Ala-Nissila}}, \ and\
  \bibinfo {author} {\bibfnamefont {S.-C.}\ \bibnamefont {Ying}},\ }\href
  {\doibase 10.1103/physreve.75.061912} {\bibfield  {journal} {\bibinfo
  {journal} {Physical Review E}\ }\textbf {\bibinfo {volume} {75}},\ \bibinfo
  {pages} {061912} (\bibinfo {year} {2007})}\BibitemShut {NoStop}%
\bibitem [{\citenamefont {Panja}\ and\ \citenamefont
  {Barkema}(2008)}]{Panja2008}%
  \BibitemOpen
  \bibfield  {author} {\bibinfo {author} {\bibfnamefont {D.}~\bibnamefont
  {Panja}}\ and\ \bibinfo {author} {\bibfnamefont {G.~T.}\ \bibnamefont
  {Barkema}},\ }\href {\doibase 10.1529/biophysj.107.116434} {\bibfield
  {journal} {\bibinfo  {journal} {Biophysical Journal}\ }\textbf {\bibinfo
  {volume} {94}},\ \bibinfo {pages} {1630} (\bibinfo {year}
  {2008})}\BibitemShut {NoStop}%
\bibitem [{\citenamefont {de~Haan}\ and\ \citenamefont
  {Slater}(2010)}]{deHaan2010}%
  \BibitemOpen
  \bibfield  {author} {\bibinfo {author} {\bibfnamefont {H.~W.}\ \bibnamefont
  {de~Haan}}\ and\ \bibinfo {author} {\bibfnamefont {G.~W.}\ \bibnamefont
  {Slater}},\ }\href {\doibase 10.1103/physreve.81.051802} {\bibfield
  {journal} {\bibinfo  {journal} {Physical Review E}\ }\textbf {\bibinfo
  {volume} {81}} (\bibinfo {year} {2010}),\
  10.1103/physreve.81.051802}\BibitemShut {NoStop}%
\bibitem [{\citenamefont {de~Haan}\ \emph {et~al.}(2015)\citenamefont
  {de~Haan}, \citenamefont {Sean},\ and\ \citenamefont {Slater}}]{deHaan2015}%
  \BibitemOpen
  \bibfield  {author} {\bibinfo {author} {\bibfnamefont {H.~W.}\ \bibnamefont
  {de~Haan}}, \bibinfo {author} {\bibfnamefont {D.}~\bibnamefont {Sean}}, \
  and\ \bibinfo {author} {\bibfnamefont {G.~W.}\ \bibnamefont {Slater}},\
  }\href {\doibase 10.1103/physreve.91.022601} {\bibfield  {journal} {\bibinfo
  {journal} {Physical Review E}\ }\textbf {\bibinfo {volume} {91}} (\bibinfo
  {year} {2015}),\ 10.1103/physreve.91.022601}\BibitemShut {NoStop}%
\bibitem [{\citenamefont {Menais}\ \emph {et~al.}(2016)\citenamefont {Menais},
  \citenamefont {Mossa},\ and\ \citenamefont {Buhot}}]{Menais2016}%
  \BibitemOpen
  \bibfield  {author} {\bibinfo {author} {\bibfnamefont {T.}~\bibnamefont
  {Menais}}, \bibinfo {author} {\bibfnamefont {S.}~\bibnamefont {Mossa}}, \
  and\ \bibinfo {author} {\bibfnamefont {A.}~\bibnamefont {Buhot}},\ }\href
  {\doibase 10.1038/srep38558} {\bibfield  {journal} {\bibinfo  {journal}
  {Scientific Reports}\ }\textbf {\bibinfo {volume} {6}},\ \bibinfo {pages}
  {38558} (\bibinfo {year} {2016})}\BibitemShut {NoStop}%
\bibitem [{\citenamefont {{Sarabadani}}\ \emph {et~al.}(2017)\citenamefont
  {{Sarabadani}}, \citenamefont {{Ghosh}}, \citenamefont {{Chaudhury}},\ and\
  \citenamefont {{Ala-Nissila}}}]{2017arXiv170809184S}%
  \BibitemOpen
  \bibfield  {author} {\bibinfo {author} {\bibfnamefont {J.}~\bibnamefont
  {{Sarabadani}}}, \bibinfo {author} {\bibfnamefont {B.}~\bibnamefont
  {{Ghosh}}}, \bibinfo {author} {\bibfnamefont {S.}~\bibnamefont
  {{Chaudhury}}}, \ and\ \bibinfo {author} {\bibfnamefont {T.}~\bibnamefont
  {{Ala-Nissila}}},\ }\href@noop {} {\bibfield  {journal} {\bibinfo  {journal}
  {ArXiv e-prints}\ } (\bibinfo {year} {2017})},\ \Eprint
  {http://arxiv.org/abs/1708.09184} {arXiv:1708.09184 [cond-mat.soft]}
  \BibitemShut {NoStop}%
\bibitem [{\citenamefont {Sakaue}(2007)}]{Sakaue2007}%
  \BibitemOpen
  \bibfield  {author} {\bibinfo {author} {\bibfnamefont {T.}~\bibnamefont
  {Sakaue}},\ }\href {\doibase 10.1103/physreve.76.021803} {\bibfield
  {journal} {\bibinfo  {journal} {Physical Review E}\ }\textbf {\bibinfo
  {volume} {76}},\ \bibinfo {pages} {021803} (\bibinfo {year}
  {2007})}\BibitemShut {NoStop}%
\bibitem [{\citenamefont {Sakaue}(2010)}]{Sakaue2010}%
  \BibitemOpen
  \bibfield  {author} {\bibinfo {author} {\bibfnamefont {T.}~\bibnamefont
  {Sakaue}},\ }\href {\doibase 10.1103/physreve.81.041808} {\bibfield
  {journal} {\bibinfo  {journal} {Physical Review E}\ }\textbf {\bibinfo
  {volume} {81}},\ \bibinfo {pages} {041808} (\bibinfo {year}
  {2010})}\BibitemShut {NoStop}%
\bibitem [{\citenamefont {Saito}\ and\ \citenamefont
  {Sakaue}(2011)}]{Saito2011}%
  \BibitemOpen
  \bibfield  {author} {\bibinfo {author} {\bibfnamefont {T.}~\bibnamefont
  {Saito}}\ and\ \bibinfo {author} {\bibfnamefont {T.}~\bibnamefont {Sakaue}},\
  }\href {\doibase 10.1140/epje/i2011-11135-3} {\bibfield  {journal} {\bibinfo
  {journal} {The European Physical Journal E}\ }\textbf {\bibinfo {volume}
  {34}} (\bibinfo {year} {2011}),\ 10.1140/epje/i2011-11135-3}\BibitemShut
  {NoStop}%
\bibitem [{\citenamefont {Saito}\ and\ \citenamefont
  {Sakaue}(2012)}]{Saito2012}%
  \BibitemOpen
  \bibfield  {author} {\bibinfo {author} {\bibfnamefont {T.}~\bibnamefont
  {Saito}}\ and\ \bibinfo {author} {\bibfnamefont {T.}~\bibnamefont {Sakaue}},\
  }\href {\doibase 10.1103/physreve.85.061803} {\bibfield  {journal} {\bibinfo
  {journal} {Physical Review E}\ }\textbf {\bibinfo {volume} {85}},\ \bibinfo
  {pages} {061803} (\bibinfo {year} {2012})}\BibitemShut {NoStop}%
\bibitem [{\citenamefont {Ambjörnsson}\ and\ \citenamefont
  {Metzler}(2005)}]{Ambjrnsson2005}%
  \BibitemOpen
  \bibfield  {author} {\bibinfo {author} {\bibfnamefont {T.}~\bibnamefont
  {Ambjörnsson}}\ and\ \bibinfo {author} {\bibfnamefont {R.}~\bibnamefont
  {Metzler}},\ }\href {\doibase 10.1103/physreve.72.030901} {\bibfield
  {journal} {\bibinfo  {journal} {Physical Review E}\ }\textbf {\bibinfo
  {volume} {72}},\ \bibinfo {pages} {030901} (\bibinfo {year}
  {2005})}\BibitemShut {NoStop}%
\bibitem [{\citenamefont {Ambjörnsson}\ \emph {et~al.}(2005)\citenamefont
  {Ambjörnsson}, \citenamefont {Lomholt},\ and\ \citenamefont
  {Metzler}}]{2Ambjrnsson2005}%
  \BibitemOpen
  \bibfield  {author} {\bibinfo {author} {\bibfnamefont {T.}~\bibnamefont
  {Ambjörnsson}}, \bibinfo {author} {\bibfnamefont {M.~A.}\ \bibnamefont
  {Lomholt}}, \ and\ \bibinfo {author} {\bibfnamefont {R.}~\bibnamefont
  {Metzler}},\ }\href {\doibase 10.1088/0953-8984/17/47/021} {\bibfield
  {journal} {\bibinfo  {journal} {Journal of Physics: Condensed Matter}\
  }\textbf {\bibinfo {volume} {17}},\ \bibinfo {pages} {S3945} (\bibinfo {year}
  {2005})}\BibitemShut {NoStop}%
\bibitem [{\citenamefont {Rowghanian}\ and\ \citenamefont
  {Grosberg}(2011)}]{Rowghanian2011}%
  \BibitemOpen
  \bibfield  {author} {\bibinfo {author} {\bibfnamefont {P.}~\bibnamefont
  {Rowghanian}}\ and\ \bibinfo {author} {\bibfnamefont {A.~Y.}\ \bibnamefont
  {Grosberg}},\ }\href {\doibase 10.1021/jp204014r} {\bibfield  {journal}
  {\bibinfo  {journal} {The Journal of Physical Chemistry B}\ }\textbf
  {\bibinfo {volume} {115}},\ \bibinfo {pages} {14127} (\bibinfo {year}
  {2011})}\BibitemShut {NoStop}%
\bibitem [{\citenamefont {Dubbeldam}\ \emph {et~al.}(2012)\citenamefont
  {Dubbeldam}, \citenamefont {Rostiashvili}, \citenamefont {Milchev},\ and\
  \citenamefont {Vilgis}}]{Dubbeldam2012}%
  \BibitemOpen
  \bibfield  {author} {\bibinfo {author} {\bibfnamefont {J.~L.~A.}\
  \bibnamefont {Dubbeldam}}, \bibinfo {author} {\bibfnamefont {V.~G.}\
  \bibnamefont {Rostiashvili}}, \bibinfo {author} {\bibfnamefont
  {A.}~\bibnamefont {Milchev}}, \ and\ \bibinfo {author} {\bibfnamefont
  {T.~A.}\ \bibnamefont {Vilgis}},\ }\href {\doibase
  10.1103/physreve.85.041801} {\bibfield  {journal} {\bibinfo  {journal}
  {Physical Review E}\ }\textbf {\bibinfo {volume} {85}},\ \bibinfo {pages}
  {041801} (\bibinfo {year} {2012})}\BibitemShut {NoStop}%
\bibitem [{\citenamefont {Ikonen}\ \emph
  {et~al.}(2012{\natexlab{a}})\citenamefont {Ikonen}, \citenamefont
  {Bhattacharya}, \citenamefont {Ala-Nissila},\ and\ \citenamefont
  {Sung}}]{Ikonen2012}%
  \BibitemOpen
  \bibfield  {author} {\bibinfo {author} {\bibfnamefont {T.}~\bibnamefont
  {Ikonen}}, \bibinfo {author} {\bibfnamefont {A.}~\bibnamefont
  {Bhattacharya}}, \bibinfo {author} {\bibfnamefont {T.}~\bibnamefont
  {Ala-Nissila}}, \ and\ \bibinfo {author} {\bibfnamefont {W.}~\bibnamefont
  {Sung}},\ }\href {\doibase 10.1103/physreve.85.051803} {\bibfield  {journal}
  {\bibinfo  {journal} {Physical Review E}\ }\textbf {\bibinfo {volume} {85}},\
  \bibinfo {pages} {051803} (\bibinfo {year} {2012}{\natexlab{a}})}\BibitemShut
  {NoStop}%
\bibitem [{\citenamefont {Ikonen}\ \emph
  {et~al.}(2012{\natexlab{b}})\citenamefont {Ikonen}, \citenamefont
  {Bhattacharya}, \citenamefont {Ala-Nissila},\ and\ \citenamefont
  {Sung}}]{2Ikonen2012}%
  \BibitemOpen
  \bibfield  {author} {\bibinfo {author} {\bibfnamefont {T.}~\bibnamefont
  {Ikonen}}, \bibinfo {author} {\bibfnamefont {A.}~\bibnamefont
  {Bhattacharya}}, \bibinfo {author} {\bibfnamefont {T.}~\bibnamefont
  {Ala-Nissila}}, \ and\ \bibinfo {author} {\bibfnamefont {W.}~\bibnamefont
  {Sung}},\ }\href {\doibase 10.1063/1.4742188} {\bibfield  {journal} {\bibinfo
   {journal} {The Journal of Chemical Physics}\ }\textbf {\bibinfo {volume}
  {137}},\ \bibinfo {pages} {085101} (\bibinfo {year}
  {2012}{\natexlab{b}})}\BibitemShut {NoStop}%
\bibitem [{\citenamefont {Ikonen}\ \emph {et~al.}(2013)\citenamefont {Ikonen},
  \citenamefont {Bhattacharya}, \citenamefont {Ala-Nissila},\ and\
  \citenamefont {Sung}}]{Ikonen2013}%
  \BibitemOpen
  \bibfield  {author} {\bibinfo {author} {\bibfnamefont {T.}~\bibnamefont
  {Ikonen}}, \bibinfo {author} {\bibfnamefont {A.}~\bibnamefont
  {Bhattacharya}}, \bibinfo {author} {\bibfnamefont {T.}~\bibnamefont
  {Ala-Nissila}}, \ and\ \bibinfo {author} {\bibfnamefont {W.}~\bibnamefont
  {Sung}},\ }\href {\doibase 10.1209/0295-5075/103/38001} {\bibfield  {journal}
  {\bibinfo  {journal} {{EPL} (Europhysics Letters)}\ }\textbf {\bibinfo
  {volume} {103}},\ \bibinfo {pages} {38001} (\bibinfo {year}
  {2013})}\BibitemShut {NoStop}%
\bibitem [{\citenamefont {Milchev}(2011)}]{Milchev2011}%
  \BibitemOpen
  \bibfield  {author} {\bibinfo {author} {\bibfnamefont {A.}~\bibnamefont
  {Milchev}},\ }\href {http://stacks.iop.org/0953-8984/23/i=10/a=103101}
  {\bibfield  {journal} {\bibinfo  {journal} {Journal of Physics: Condensed
  Matter}\ }\textbf {\bibinfo {volume} {23}},\ \bibinfo {pages} {103101}
  (\bibinfo {year} {2011})}\BibitemShut {NoStop}%
\bibitem [{\citenamefont {Luo}\ \emph {et~al.}(2009)\citenamefont {Luo},
  \citenamefont {Ala-Nissila}, \citenamefont {Ying},\ and\ \citenamefont
  {Metzler}}]{Luo2009}%
  \BibitemOpen
  \bibfield  {author} {\bibinfo {author} {\bibfnamefont {K.}~\bibnamefont
  {Luo}}, \bibinfo {author} {\bibfnamefont {T.}~\bibnamefont {Ala-Nissila}},
  \bibinfo {author} {\bibfnamefont {S.-C.}\ \bibnamefont {Ying}}, \ and\
  \bibinfo {author} {\bibfnamefont {R.}~\bibnamefont {Metzler}},\ }\href
  {\doibase 10.1209/0295-5075/88/68006} {\bibfield  {journal} {\bibinfo
  {journal} {{EPL} (Europhysics Letters)}\ }\textbf {\bibinfo {volume} {88}},\
  \bibinfo {pages} {68006} (\bibinfo {year} {2009})}\BibitemShut {NoStop}%
\bibitem [{Note1()}]{Note1}%
  \BibitemOpen
  \bibinfo {note} {See Supplemental Material at [URL will be inserted by
  publisher] for supporting text and figures}\BibitemShut {NoStop}%
\bibitem [{\citenamefont {Linak}\ \emph {et~al.}(2011)\citenamefont {Linak},
  \citenamefont {Tourdot},\ and\ \citenamefont {Dorfman}}]{Linak2011}%
  \BibitemOpen
  \bibfield  {author} {\bibinfo {author} {\bibfnamefont {M.~C.}\ \bibnamefont
  {Linak}}, \bibinfo {author} {\bibfnamefont {R.}~\bibnamefont {Tourdot}}, \
  and\ \bibinfo {author} {\bibfnamefont {K.~D.}\ \bibnamefont {Dorfman}},\
  }\href {\doibase 10.1063/1.3662137} {\bibfield  {journal} {\bibinfo
  {journal} {J. Chem. Phys.}\ }\textbf {\bibinfo {volume} {135}},\ \bibinfo
  {eid} {205102} (\bibinfo {year} {2011})}\BibitemShut {NoStop}%
\bibitem [{\citenamefont {Bell}\ \emph {et~al.}(2012)\citenamefont {Bell},
  \citenamefont {Engst}, \citenamefont {Ablay}, \citenamefont {Divitini},
  \citenamefont {Ducati}, \citenamefont {Liedl},\ and\ \citenamefont
  {Keyser}}]{Bell2012}%
  \BibitemOpen
  \bibfield  {author} {\bibinfo {author} {\bibfnamefont {N.~A.~W.}\
  \bibnamefont {Bell}}, \bibinfo {author} {\bibfnamefont {C.~R.}\ \bibnamefont
  {Engst}}, \bibinfo {author} {\bibfnamefont {M.}~\bibnamefont {Ablay}},
  \bibinfo {author} {\bibfnamefont {G.}~\bibnamefont {Divitini}}, \bibinfo
  {author} {\bibfnamefont {C.}~\bibnamefont {Ducati}}, \bibinfo {author}
  {\bibfnamefont {T.}~\bibnamefont {Liedl}}, \ and\ \bibinfo {author}
  {\bibfnamefont {U.~F.}\ \bibnamefont {Keyser}},\ }\href {\doibase
  10.1021/nl204098n} {\bibfield  {journal} {\bibinfo  {journal} {Nano Letters}\
  }\textbf {\bibinfo {volume} {12}},\ \bibinfo {pages} {512} (\bibinfo {year}
  {2012})}\BibitemShut {NoStop}%
\bibitem [{\citenamefont {Hernandez-Ainsa}\ \emph {et~al.}(2013)\citenamefont
  {Hernandez-Ainsa}, \citenamefont {Bell}, \citenamefont {Thacker},
  \citenamefont {Göpfrich}, \citenamefont {Misiunas}, \citenamefont
  {Fuentes-Perez}, \citenamefont {Moreno-Herrero},\ and\ \citenamefont
  {Keyser}}]{HernndezAinsa2013}%
  \BibitemOpen
  \bibfield  {author} {\bibinfo {author} {\bibfnamefont {S.}~\bibnamefont
  {Hernandez-Ainsa}}, \bibinfo {author} {\bibfnamefont {N.~A.~W.}\ \bibnamefont
  {Bell}}, \bibinfo {author} {\bibfnamefont {V.~V.}\ \bibnamefont {Thacker}},
  \bibinfo {author} {\bibfnamefont {K.}~\bibnamefont {Göpfrich}}, \bibinfo
  {author} {\bibfnamefont {K.}~\bibnamefont {Misiunas}}, \bibinfo {author}
  {\bibfnamefont {M.~E.}\ \bibnamefont {Fuentes-Perez}}, \bibinfo {author}
  {\bibfnamefont {F.}~\bibnamefont {Moreno-Herrero}}, \ and\ \bibinfo {author}
  {\bibfnamefont {U.~F.}\ \bibnamefont {Keyser}},\ }\href {\doibase
  10.1021/nn401759r} {\bibfield  {journal} {\bibinfo  {journal} {{ACS} Nano}\
  }\textbf {\bibinfo {volume} {7}},\ \bibinfo {pages} {6024} (\bibinfo {year}
  {2013})}\BibitemShut {NoStop}%
\bibitem [{\citenamefont {Trepagnier}\ \emph {et~al.}(2007)\citenamefont
  {Trepagnier}, \citenamefont {Radenovic}, \citenamefont {Sivak}, \citenamefont
  {Geissler},\ and\ \citenamefont {Liphardt}}]{Trepagnier2007}%
  \BibitemOpen
  \bibfield  {author} {\bibinfo {author} {\bibfnamefont {E.~H.}\ \bibnamefont
  {Trepagnier}}, \bibinfo {author} {\bibfnamefont {A.}~\bibnamefont
  {Radenovic}}, \bibinfo {author} {\bibfnamefont {D.}~\bibnamefont {Sivak}},
  \bibinfo {author} {\bibfnamefont {P.}~\bibnamefont {Geissler}}, \ and\
  \bibinfo {author} {\bibfnamefont {J.}~\bibnamefont {Liphardt}},\ }\href
  {\doibase 10.1021/nl0714334} {\bibfield  {journal} {\bibinfo  {journal} {Nano
  Letters}\ }\textbf {\bibinfo {volume} {7}},\ \bibinfo {pages} {2824}
  (\bibinfo {year} {2007})}\BibitemShut {NoStop}%
\bibitem [{\citenamefont {Humphrey}\ \emph {et~al.}(1996)\citenamefont
  {Humphrey}, \citenamefont {Dalke},\ and\ \citenamefont
  {Schulten}}]{Humphrey1996}%
  \BibitemOpen
  \bibfield  {author} {\bibinfo {author} {\bibfnamefont {W.}~\bibnamefont
  {Humphrey}}, \bibinfo {author} {\bibfnamefont {A.}~\bibnamefont {Dalke}}, \
  and\ \bibinfo {author} {\bibfnamefont {K.}~\bibnamefont {Schulten}},\ }\href
  {http://www.ks.uiuc.edu/Research/vmd/} {\bibfield  {journal} {\bibinfo
  {journal} {J. Mol. Graph.}\ }\textbf {\bibinfo {volume} {14}},\ \bibinfo
  {pages} {33} (\bibinfo {year} {1996})}\BibitemShut {NoStop}%
\bibitem [{\citenamefont {Plimpton}(1995)}]{Plimpton1995}%
  \BibitemOpen
  \bibfield  {author} {\bibinfo {author} {\bibfnamefont {S.}~\bibnamefont
  {Plimpton}},\ }\href {\doibase 10.1006/jcph.1995.1039} {\bibfield  {journal}
  {\bibinfo  {journal} {J. Comp. Phys.}\ }\textbf {\bibinfo {volume} {117}},\
  \bibinfo {pages} {1} (\bibinfo {year} {1995})}\BibitemShut {NoStop}%
\bibitem [{\citenamefont {{Suhonen}}\ \emph {et~al.}(2017)\citenamefont
  {{Suhonen}}, \citenamefont {{Piili}},\ and\ \citenamefont
  {{Linna}}}]{2017arXiv170706663S}%
  \BibitemOpen
  \bibfield  {author} {\bibinfo {author} {\bibfnamefont {P.~M.}\ \bibnamefont
  {{Suhonen}}}, \bibinfo {author} {\bibfnamefont {J.}~\bibnamefont {{Piili}}},
  \ and\ \bibinfo {author} {\bibfnamefont {R.~P.}\ \bibnamefont {{Linna}}},\
  }\href@noop {} {\bibfield  {journal} {\bibinfo  {journal} {ArXiv e-prints}\ }
  (\bibinfo {year} {2017})},\ \Eprint {http://arxiv.org/abs/1707.06663}
  {arXiv:1707.06663 [physics.bio-ph]} \BibitemShut {NoStop}%
\bibitem [{\citenamefont {Ling}\ and\ \citenamefont {Ling}(2013)}]{Ling2013}%
  \BibitemOpen
  \bibfield  {author} {\bibinfo {author} {\bibfnamefont {D.~Y.}\ \bibnamefont
  {Ling}}\ and\ \bibinfo {author} {\bibfnamefont {X.~S.}\ \bibnamefont
  {Ling}},\ }\href {\doibase 10.1088/0953-8984/25/37/375102} {\bibfield
  {journal} {\bibinfo  {journal} {Journal of Physics: Condensed Matter}\
  }\textbf {\bibinfo {volume} {25}},\ \bibinfo {pages} {375102} (\bibinfo
  {year} {2013})}\BibitemShut {NoStop}%
\bibitem [{\citenamefont {Sarabadani}\ \emph {et~al.}(2017)\citenamefont
  {Sarabadani}, \citenamefont {Ikonen}, \citenamefont {M\"{o}kk\"{o}nen},
  \citenamefont {Ala-Nissila}, \citenamefont {Carson},\ and\ \citenamefont
  {Wanunu}}]{Sarabadani2017}%
  \BibitemOpen
  \bibfield  {author} {\bibinfo {author} {\bibfnamefont {J.}~\bibnamefont
  {Sarabadani}}, \bibinfo {author} {\bibfnamefont {T.}~\bibnamefont {Ikonen}},
  \bibinfo {author} {\bibfnamefont {H.}~\bibnamefont {M\"{o}kk\"{o}nen}},
  \bibinfo {author} {\bibfnamefont {T.}~\bibnamefont {Ala-Nissila}}, \bibinfo
  {author} {\bibfnamefont {S.}~\bibnamefont {Carson}}, \ and\ \bibinfo {author}
  {\bibfnamefont {M.}~\bibnamefont {Wanunu}},\ }\href {\doibase
  10.1038/s41598-017-07227-3} {\bibfield  {journal} {\bibinfo  {journal}
  {Scientific Reports}\ }\textbf {\bibinfo {volume} {7}} (\bibinfo {year}
  {2017}),\ 10.1038/s41598-017-07227-3}\BibitemShut {NoStop}%
\bibitem [{\citenamefont {Sakaue}\ \emph {et~al.}(2012)\citenamefont {Sakaue},
  \citenamefont {Saito},\ and\ \citenamefont {Wada}}]{Sakaue2012}%
  \BibitemOpen
  \bibfield  {author} {\bibinfo {author} {\bibfnamefont {T.}~\bibnamefont
  {Sakaue}}, \bibinfo {author} {\bibfnamefont {T.}~\bibnamefont {Saito}}, \
  and\ \bibinfo {author} {\bibfnamefont {H.}~\bibnamefont {Wada}},\ }\href
  {\doibase 10.1103/physreve.86.011804} {\bibfield  {journal} {\bibinfo
  {journal} {Physical Review E}\ }\textbf {\bibinfo {volume} {86}} (\bibinfo
  {year} {2012}),\ 10.1103/physreve.86.011804}\BibitemShut {NoStop}%
\bibitem [{\citenamefont {Bacci}\ \emph {et~al.}(2012)\citenamefont {Bacci},
  \citenamefont {Chinappi}, \citenamefont {Casciola},\ and\ \citenamefont
  {Cecconi}}]{Bacci2012}%
  \BibitemOpen
  \bibfield  {author} {\bibinfo {author} {\bibfnamefont {M.}~\bibnamefont
  {Bacci}}, \bibinfo {author} {\bibfnamefont {M.}~\bibnamefont {Chinappi}},
  \bibinfo {author} {\bibfnamefont {C.~M.}\ \bibnamefont {Casciola}}, \ and\
  \bibinfo {author} {\bibfnamefont {F.}~\bibnamefont {Cecconi}},\ }\href
  {\doibase 10.1021/jp300143x} {\bibfield  {journal} {\bibinfo  {journal} {The
  Journal of Physical Chemistry B}\ }\textbf {\bibinfo {volume} {116}},\
  \bibinfo {pages} {4255} (\bibinfo {year} {2012})}\BibitemShut {NoStop}%
\bibitem [{\citenamefont {Vocks}\ \emph {et~al.}(2009)\citenamefont {Vocks},
  \citenamefont {Panja},\ and\ \citenamefont {Barkema}}]{Vocks2009}%
  \BibitemOpen
  \bibfield  {author} {\bibinfo {author} {\bibfnamefont {H.}~\bibnamefont
  {Vocks}}, \bibinfo {author} {\bibfnamefont {D.}~\bibnamefont {Panja}}, \ and\
  \bibinfo {author} {\bibfnamefont {G.~T.}\ \bibnamefont {Barkema}},\ }\href
  {\doibase 10.1088/0953-8984/21/37/375105} {\bibfield  {journal} {\bibinfo
  {journal} {Journal of Physics: Condensed Matter}\ }\textbf {\bibinfo {volume}
  {21}},\ \bibinfo {pages} {375105} (\bibinfo {year} {2009})}\BibitemShut
  {NoStop}%
\bibitem [{\citenamefont {Szymczak}(2016)}]{Szymczak2016}%
  \BibitemOpen
  \bibfield  {author} {\bibinfo {author} {\bibfnamefont {P.}~\bibnamefont
  {Szymczak}},\ }\href {\doibase 10.1038/srep21702} {\bibfield  {journal}
  {\bibinfo  {journal} {Sci. Rep.}\ }\textbf {\bibinfo {volume} {6}},\ \bibinfo
  {pages} {21702} (\bibinfo {year} {2016})}\BibitemShut {NoStop}%
\end{thebibliography}%

\end{document}